\documentclass[pre,twocolumn,superscriptaddress,floatfix]{revtex4}
\usepackage{epsfig}
\usepackage{float}
\usepackage{graphicx}
\usepackage{amssymb}
\usepackage{mathtools}
\usepackage{color}

\begin{document}

\title{Crystallization of Lennard-Jones nanodroplets: from near melting to deeply supercooled}

\author{Shahrazad M.~A.~Malek}
\affiliation{Department of Physics and Physical Oceanography,
Memorial University of Newfoundland, St. John's, NL, A1B 3X7, Canada}

\author{Gregory P. Morrow}
\affiliation{Department of Physics and Physical Oceanography,
Memorial University of Newfoundland, St. John's, NL, A1B 3X7, Canada}

\author{Ivan Saika-Voivod}
\affiliation{Department of Physics and Physical Oceanography,
Memorial University of Newfoundland, St. John's, NL, A1B 3X7, Canada}

\date{\today}

\begin{abstract}

We carry out molecular dynamics (MD) and Monte Carlo (MC) simulations to characterize nucleation in liquid clusters 
of 600 Lennard-Jones particles over a broad range of temperatures.  
We use the formalism of mean first-passage times to determine the rate and find that Classical Nucleation Theory (CNT)
predicts the rate quite well, even when employing simple modelling of crystallite shape, chemical potential, surface tension
and particle attachment rate, 
down to the temperature where the droplet loses metastability and crystallization proceeds through growth-limited nucleation
in an unequilibrated liquid. 
Below this crossover temperature, the nucleation rate is still predicted when MC simulations are used to directly calculate quantities required by CNT.
Discrepancy in critical embryo sizes obtained from MD and MC arises 
when twinned structures with five-fold symmetry provide a competing free energy pathway out of the critical region.
We find that crystallization begins with hcp-fcc stacked precritical nuclei and differentiation to various end structures occurs
when these embryos become critical.
We confirm that using the largest embryo in the system as a reaction coordinate is useful in determining the onset
of growth-limited nucleation and show that it gives the same free energy barriers as the full cluster size distribution once the proper reference state is identified.
We find that the bulk melting temperature controls the rate, even though the solid-liquid coexistence temperature
for the droplet is significantly lower.  
The value of surface tension that renders close agreement between CNT and direct rate determination is 
significantly lower than what is expected for the bulk system.

\end{abstract}

\maketitle

\section{Introduction}

Nanotechnology has garnered much interest in the last few decades because of the wide range of applications that come out of it. Nanoclusters, small clusters 
comprising tens to millions of atoms, are used in a variety of settings, such as tuning the optical~\cite{Puzder,Ossicini,Townsend} and electronic properties of materials~\cite{Ossicini,Kruppa}, biolabeling and 
imaging~\cite{Venkatesh}, catalysis~\cite{Visikovskiy,Mitsuhara}, and chemical sensing~\cite{Penza}.  The various structures to which nanoclusters solidify, as well as their surface properties, bear a strong impact
on their function~\cite{baletto}.

Much attention has been paid to the size dependence of nanocluster structure.  
Experimental work on argon clusters showed that for fewer than 50 atoms, polyicosahedral structure emerges~\cite{Torchet-83}, for larger particles up to 750 atoms
multilayer icosahedra are formed, while beyond this size the structure becomes fcc~\cite{Torchet-86}.  Simulations with the Lennard-Jones (LJ) potential,
a reasonable model for noble gases, as well as exhaustive searches of ground state structures confirmed this picture~\cite{Torchet-89,WALEDOYE, WALESBOOK}.
LJ simulations generally reveal rather rich behavior, especially at finite temperature $T$, in terms of local and global structures, 
transformations, size dependence and role of the surface~\cite{JonathanWales, polak2003, Xueguang, polak2006,noyadoye, Mandelshtam, polak2008,polak2013a,polak2013b,polak2014}.
Our interest is how various structures form out of the liquid state on cooling.

Freezing of a liquid generally occurs through the process of nucleation.  This is accomplished when one of the embryonic crystallites
that appear as structural fluctuations in the liquid reaches a sufficient size to overcome the crystal-liquid surface tension that tends to 
shrink and eliminate small crystalline embryos.  Classical Nucleation Theory (CNT) forms the basis of understanding the process qualitatively and provides
quantitative predictions for the rate of nucleation.  Central to CNT is $\Delta G(n)$, 
the reversible work required to form an embryo of size $n$ particles of the stable
phase within the metastable bulk~\cite{Debenedetti}.  However, the predicted rate is highly sensitive to this work,
and therefore to such considerations as the shape of the embryos,
the nature of the interface and to the potentially $T$ and curvature dependent surface tension.

The freezing of nanodroplets, i.e., nanoclusters in their liquid form, is complicated by the fact that such small
systems can often freeze into more than one structure, for example icosahaderal, decahedral or bulk-like fcc and hcp structures.
And hence the nucleation process is potentially competitive in nanodroplets~\cite{RichardMolecular}.
One wonders at what point during the freezing
process does differentiation between structures occur and whether CNT provides a reasonable description of the rate at all.  These are unresolved 
questions and their answers are likely system specific.  

One study employing simulations of gold nanoparticles found that at sufficient supercooling, CNT predicted a constant or decreasing freezing rate with
further supercooling while direct simulations saw the reverse, namely an increasing rate with further cooling~\cite{eduardo}.
This peculiar result is connected to broader questions regarding the choice of reaction coordinate in describing the nucleation process
and the resulting free energy landscape, the description of nucleation when barriers are low and the approach to a possible spinodal-like
end to liquid metastability~\cite{spinodal}.  Spinodal-like nucleation has been suggested to occur for bulk LJ~\cite{trudu}, but this idea has 
been challenged~\cite{wangklein,bartell}.

In the present study, we use molecular dynamics (MD) simulations to determine the freezing rate of a droplet consisting of 
600 LJ particles.
We press into service the mean first-passage time (MFPT) formalism of Reguera and 
co-workers~\cite{RegueraMFPT, RegueraKinetic, RegueraCROSS} to determine the rate
and critical cluster sizes over a broad range of $T$.  
While generally for nanodroplets the surface may play a large role
in determining the rate, since a large fraction of particles is near or at the surface of the droplet, crystallization for the present system occurs
within the interior~\cite{TheRoleOfFcc}.  We thus expect CNT as formulated for bulk liquids to hold without the modifications often employed to describe nucleation occurring on the surface~\cite{bowlescap-NR}. 

The previous study of this system~\cite{TheRoleOfFcc} also revealed that several competing structures, some based on fcc tetrahedra of different sizes, exist as basins within the
free energy landscape of the system.  However, as the free energy was calculated as a function of global measures of surface 
and bulk crystallinity, little light was shed on the question of how these different structures arise.  


This paper is organized as follows.  In Section II, we review and discuss some aspects of CNT and MFPT, 
while we provide details of our simulations in Section III.  We report our results in Section IV, including
a determination of the liquid-solid coexistence temperature, the freezing rate as a function of $T$ from MD simulations,
modelling the $T$ dependence of the rate through CNT, determining the free energy of crystallite formation, and 
an analysis of critical nuclei structure.
Section V provides a discussion of our results before summarizing our conclusions in Section VI.

\section{CNT, MFPT and the low barrier regime}

\subsection{CNT}

According to CNT~\cite{Debenedetti,Das}, the rate of nucleation $J$, that is to say the number of crystalline embryos that cross 
the critical size threshold and start to grow per unit time in the steady-state,
is given by,
\begin{equation}\label{eqCNTrate}
J_{\rm CNT} = N_p Z f_{\rm crit}^+ \exp{\left(-\beta \Delta G^*\right)},
\end{equation}
where $N_p$ is the number of molecules in the system, the Zeldovich factor is 
$Z= \left( \frac{\beta}{2 \pi} \left | \frac{\partial^2 \Delta G(n^*)}{\partial n^2}\right |  \right)^{1/2}$,
$\beta = (k_{\rm B} T)^{-1}$ with $k_{\rm B}$ the Boltzmann constant,
$\Delta G^* = \Delta G(n^*)$, the minimum work required to form an embryo of critical size
$n^*$, and
$f_{\rm crit}^+$ is the attachment rate of molecules to an embryo of size $n^*$.  
We note that, at variance with Eq.~\ref{eqCNTrate}, the rate is often stated in terms of the number of nucleation events per unit time per unit
volume.  Here we have absorbed the volume of the system into $J_{\rm CNT}$.  

The reversible work required to assemble an embryo of size $n$ is related to the distribution of embryo sizes in the
system~\cite{Das},
\begin{equation}\label{eqdelg}
\beta \Delta G(n) = -\ln \left[ \frac{N(n)}{N_t} \right] \approx -\ln \left[ \frac{N(n)}{N_p} \right],
\end{equation}
where $N(n)$ is the equilibrium number of embryos of size $n$ in the system and 
$N_t = \sum_{i=0}^h N(i)$ is the total number of embryos (including liquid-like particles) in the system and is approximated by $N_p$
since we assume that the system is dominated by liquid-like particles, and $h$ is a constraint on the largest embryo size
that is necessary to formally define the metastable equilibrium state.  $N(0)$ refers to the number of liquid-like particles in the system, while
$N(1)$ refers to the number of particles that are themselves crystal-like, but the neighbours of which are liquid-like.   
Because of surface tension $\gamma$ between liquid and crystal, 
$\Delta G(n)$ is initially positively sloped and possesses a maximum at $n^*$.


The simplest model for the work of crystallite formation is~\cite{Debenedetti,Das},
\begin{equation}
\beta\Delta G(n) = -\beta \Delta\mu \,n + \beta\gamma A \,n^{2/3}, 
\label{eqwork-min}
\end{equation}
where $\Delta\mu=\mu_L-\mu_S$ is the difference between the chemical potentials for the bulk phase $\mu_L$ and the embryo phase $\mu_S$, with $\Delta\mu>0$,  and $A$ is a shape-dependent proportionality constant that assumes that embryos are compact, i.e., for an embryo of volume  $\sim n$, the surface area should be  $S = A n^{2/3}$. For spherical embryos, $A=\sqrt[3]{36\pi v^2}$, where $v$ is the volume per particle in the embryonic phase.  Within this model,
$\beta \Delta G^* = \frac{4}{27} \frac{\left(\beta A \gamma\right)^3}{\left(\beta \Delta\mu\right)^2}$,
$n^* = \frac{8}{27} \frac{\left( \beta A\gamma\right)^3}{\left(\beta \Delta\mu\right)^3} $ and
$Z = \frac{3}{4 \sqrt{\pi}} \frac{\left(\beta \Delta\mu\right)^2}{\left(\beta A \gamma\right)^{3/2}}$.

The simplest model for the $T$-dependence of $J_{\rm CNT}$  is obtained by combining Eqs.~\ref{eqCNTrate} and~\ref{eqwork-min}, along with assuming $\gamma$ and $A$ constant.  By further assuming a constant enthalpy difference $\Delta H$ between the liquid and crystal phases as $T$ decreases at constant pressure $p$, one obtains,
\begin{equation}\label{eqbetadelmu}
\beta \Delta\mu = \frac{\Delta H}{N_p k_{\rm B}} \frac{T_m - T}{T T_m}, 
\end{equation}
where $T_m$ is the melting temperature of the bulk crystal, and at which point $J_{\rm CNT}$ is zero.  
Additionally, one assumes a
simple Arrhenius $T$ dependence of the critical attachment rate,
\begin{equation}\label{eqfcritT}
f_{\rm crit}^+ = f_0 \exp{\left(-\frac{C}{T} \right)},
\end{equation}
where $k_{\rm B} C$ is an activation free energy.
Combining all these approximations results in~\cite{okui,clouet-NR},
\begin{equation}\label{eqstar} 
J_{\rm CNT}(T)= \lambda\frac{{(T_m-T)}^2}{\sqrt{T}}\exp\left[-\frac{C}{T}-\frac{B}{T{(T_m-T)}^2}\right],
\end{equation}
which predicts a maximum rate to occur even in the absence of considerable slowing down of dynamics.
The simple modelling employed implies that the barrier to nucleation
is,
\begin{equation}
\beta \Delta G^*  = \frac{B}{T ( T_m - T)^2}, 
\label{eqGstarTCNT}
\end{equation}
and therefore has a minimum at $T_m/3$, which tends to maximize the rate, before 
diverging as $T$ approaches zero.  
In terms of the physical quantities $\Delta H$, $T_m$, $f_0$, $A$ and $\gamma$, the parameters $\lambda$ and $B$ in the model are given by,
\begin{equation}\label{eqlambda}
\lambda = f_0 N_p \frac{3 }{4 \sqrt{\pi k_{\rm B}}} \frac{ 1}{\left(  A \gamma \right)^{3/2}} \left( \frac{\Delta H}{N_p} \right)^2 \frac{1}{T_m^2},
\end{equation}
\begin{equation}\label{eqB}
B = \frac{4}{27}   \frac{\left(A \gamma\right)^3}{k_{\rm B}}  \left( \frac{N_p}{\Delta H} \right)^2 T_m^2.
\end{equation}
The quantities $\lambda$, $T_m$, $B$ and $C$ can, in principal, be obtained through fitting the rate
as a function of $T$ with Eq.~\ref{eqstar}.

\subsection{$n_{\rm max}$ as the order parameter}

In the present work, as is now common in simulation studies of nucleation, we employ the size of the largest embryo in the system $n_{\rm max}$ as 
a reaction coordinate.  Once an embryo definition is set, every system configuration can be uniquely assigned a value of $n_{\rm max}$, and
hence the (configurational part) 
of a restricted partition function can be defined through~\cite{eduardo},
\begin{equation}
Q(n_{\rm max}) = \sum_{c \in n_{\rm max}} \exp{\left( -\beta U_c \right)},
\end{equation}
where $U_c$ is the potential energy of configuration $c$, 
restricted to those configurations that have a largest embryo of size $n_{\rm max}$.
We can then further define the free energy~\cite{ben},
\begin{equation}\label{eqdelFq}
\beta \Delta F(n) =  -\ln{\left[ \frac{Q(n)}{Q_{\rm liq}}\right]},
\end{equation}
where we have dropped the subscript on $n$ for notational convenience and 
$Q_{\rm liq}$ is the partition function of the metastable liquid, defined as,
\begin{equation}\label{eqQnliq}
Q_{\rm liq} = \sum_{n=0}^{n_F^*} Q(n),
\end{equation}
where $n_F^*$ is the (critical) cluster size at which 
$\beta \Delta F(n)$ possesses a local maximum, i.e., where $Q(n)$ has a local minimum.  
So defined, $\beta \Delta F(n)$ is directly related to the probability that the largest cluster in 
the system is of size $n$, given that the system is in the metastable liquid, $P_{\rm max}(n)$,
\begin{equation}\label{eqdelF}
\beta \Delta F(n) = -\ln{ P_{\rm max}(n)}.
\end{equation}
That is, the normalization is such that,
\begin{equation}\label{eqpmaxnorm}
\sum_{n=0}^{n_F^*} P_{\rm max}(n) =1.
\end{equation}

For relatively large barrier heights, large embryos are rare, i.e., there is only one large embryo in the system if there is one at all.
This implies the equality of the following three quantities: the probability of there being an embryo of size $n$ in the system; the probability that the largest embryo is of size $n$;
and the average number of embryos of size $n$.  This becomes immediately obvious when constructing related  
histograms during  the simulations.  In this regime, $P_{\rm max}(n) = N(n)$ (and both are small).  

The TST rate expression when there is a free energy barrier present is,
\begin{equation}\label{eqTSTrate}
J_{\rm TST} = f^+(n_F^*) Z_F \exp{\left[-\beta \Delta F^*\right]},
\end{equation}
where $n_F^*$, the Zeldovich factor $Z_F = \left [ \beta \Delta F''(n_F^*)/(2 \pi)\right]^{1/2}$ and $f^+(n_F^*)$, the generalized diffusion coefficient at the critical state, 
become equal to $n^*$,  $Z$ and $f_{\rm crit}^+$ at sufficiently high barriers, respectively, and $\beta \Delta F^* = \beta \Delta F(n_F^*)$.  
$f_{\rm crit}^+$ in Eq.~\ref{eqCNTrate} is the attachment rate 
of particles to an embryo of critical size, while $f^+(n_F^*)$ tracks changes in the size of the largest embryo at critical size in the system.  The two are the
same so long as the largest embryo in the system is the only embryo near the critical size.  Again, when barriers are high, the equalities $n^*=n_F^*$ and 
$P_{\rm max}(n) = N(n)$ near $n^*$ imply that  $\beta \Delta G^* = \beta \Delta F^* + \ln{N_p}$, and this is consistent
when comparing Eqs.~\ref{eqCNTrate} and \ref{eqTSTrate}.  However, there is no reason why this should hold when barriers become low.

It is generally the case that $\Delta F(n)$ possesses a minimum at $n_{\rm min}$, the most likely largest embryo size.
It is tempting to formulate Eq.~\ref{eqTSTrate} in terms of the free energy difference,
\begin{equation}
\beta\Delta F^*_{\rm min} = -\ln{\left[\frac{ P_{\rm max}(n_F^*)}{P_{\rm max}(n_{\rm min})}\right] = \beta \Delta F^* - \beta \Delta F(n_{\rm min})}.
\end{equation}
This is incorrect in terms of rate prediction,  as it fails to account for the phase space available in the free energy
basin around $n_{\rm min}$~\cite{ben}.

The identification of $\Delta F^*_{\rm min} \rightarrow 0$ with a spinodal has been shown to be incorrect~\cite{wangklein}, but it nonetheless marks the point at which
the liquid system ceases to possess a basin in the free energy and has therefore lost formal metastability.  For bulk systems of finite size, this marks the point at which 
phase change proceeds through the monotonic increase in size of the largest embryo in the system with time, i.e., because the system is large enough, it becomes
probable that it possesses an embryo of critical size as soon as diffusive particle attachment  allows.  Phase transformation of the sample thus proceeds through growth-limited nucleation~\cite{RegueraCROSS}.  However, the metastable
phase has not lost inherent metastability as work is still required to form an embryo.  For systems such a our nanodroplets, it is perhaps not meaningful to distinguish between phase and system, but we
nonetheless expect that the loss of metastability occurring at  $\Delta F^*_{\rm min}=0$ to be actualized through a growth-limited nucleation mechanism with a transformation 
rate given, at least approximately, by Eq.~\ref{eqCNTrate}.  A true kinetic spinodal, i.e., a loss of stability on the particle level, should occur when $\Delta G^*$ vanishes.

\subsection{MFPT}

In recent times, Reguera and co-workers reformulated the use of mean first-passage time from TST~\cite{RegueraMFPT,RegueraKinetic, RegueraCROSS,TST-NR} in order to characterize 
the nucleation process in the regime where nucleation times are accessible by direct MD simulations.  In this MFPT formalism, 
when the time to crystallize is dominated by barrier crossing, the mean time at which the largest crystalline embryo in the system first reaches
size $n$ is given by,
\begin{equation}
 \tau(n) = \frac{\tau_J}{2}\{1+{\rm{erf}}[Z_F \sqrt{\pi}(n-n_F^*)]\},
\label{eqMFPT}
\end{equation}
where $\tau_J = 1/J$.  Thus, calculating $\tau(n)$ from an ensemble of simulations for which crystallization takes place, yields good estimates
of $J$ as well as $Z$ and $n^*$.

Typically, as supercooling increases, the sigmoidal shape  of the MFPT becomes less well described by Eq.~\ref{eqMFPT}, and we can instead estimate $n_F^*$ through~\cite{RegueraMFPT},
\begin{equation}\label{inflection}
\frac{\partial^2 \tau(n_F^*)}{\partial n^2} \approx 0.
\end{equation}

\section{Model and simulations}

Our system consists of $N_p=600$ particles interacting through the LJ pair potential,
$U_{LJ}(r)=4\epsilon \left[\left(\frac{\sigma}{r}\right)^{12}-\left(\frac{\sigma}{r}\right)^{6}\right]$, simulated in the canonical ensemble.
All reported quantities are given in reduced dimensionless units, e.g., 
length is rescaled by $\sigma$, energy by $\epsilon$,
time by $\sqrt{\epsilon/(m \sigma^2)}$ (where $m$ is the mass of a particle), temperature by $\epsilon/k_{\rm B}$ and pressure by $\epsilon/\sigma^3$.
We use a cubic simulation box of side length $L=30$ and employ a potential cutoff of $R_c=14.99999$.  For the range of $T$ we consider,
the system consists of a single condensed droplet with a few particles at most detaching themselves from the droplet  into the surrounding gas phase.  The finite size 
and periodic boundaries ensure that these particles can return to the droplet and that the droplet does not evaporate.  The box size is sufficiently large
to ensure that particles within the droplet do not interact unphysically with periodic images of the droplet.

We use Gromacs v4.5.5~\cite{GROMACS} to carry out MD simulations.  Temperature is maintained with the Nos{\'e}-Hoover thermostat with a time constant of 1.
We use a time step of $\Delta t = 0.001$ and integrate equations of motion with the leap-frog algorithm.
We equilibrate the system at $T=0.53$, for which the droplet is well formed but clearly a liquid, and subsequently harvest 501 independent configurations 
by sampling every $100\, 000$ time steps.  Each of these configurations serves as a starting point for a ``crystallization run'', for which the thermostat is
set to the desired lower $T$.  We determine $\tau(n)$ from the MFPT formalism, as in Refs.~\cite{RegueraMFPT,SarahIvan} from these 501 crystallization trajectories for each of several
$T$ from 0.490 down to 0.385 in steps of 0.005, and from 0.350 to 0.100 in steps of 0.05.  To determine $\tau(n)$, we
calculate the size of the largest crystalline embryo, as described below, every 1000 time steps (integer LJ time units).

We employ the procedure developed by Frenkel and co-workers~\cite{FrenkelAuer,Frenkelsalt} to define crystal-like embryos within the droplet, based on quantifying the local bond ordering for a single particle via spherical harmonics~\cite{Steinhardt}.
See also Refs.~\cite{SarahIvan,romano} for details.  In this procedure, there are three parameters: the distance cutoff for determining whether two particles are neighbors, chosen from the minimum at $r=1.363$ between first and second peaks
of the radial distribution function; a threshold for the correlation $c_{\ij}$, a complex dot product that determines whether two neighboring particles have sufficiently aligned local bonding patterns and above which the particles are considered to be {\it connected}, which we choose to be $0.5$ as the intersection point for the probability distributions of $c_{ij}$ obtained from 100 liquid and 100 solidified configurations at $T=0.475$; and the number of connections a particle needs in order to be considered solid-like, which we take to be 0.8 times the number of neighbors a particles has (keeping in mind that particles on the surface have fewer neighbors).  Further, for the purposes of finding the size distribution of embryos, two connected, crystal-like particles are considered to be part of the same crystalline embryo.

In order to differentiate between embryos of the same size but different overall structure, we determine  the overall crystallinity of the cluster by calculating the often-used quantity $Q_6$~\cite{FrenkelNumericalLJ}.

To determine the free energy profiles, we
carry out umbrella sampling Monte Carlo (MC) simulations in the canonical ensemble 
to determine the works defined in Eqs.~\ref{eqdelg} and \ref{eqdelF}. 
When barriers are reasonably high, we make use of a biasing potential, 
\begin{equation}
\phi(n_{\rm max})=\frac{1}{2}\kappa (n_{\rm max} - n_0)^2,
\end{equation}
where $\kappa=0.00625$ determines the strength of the constraint and $n_0$ is the target embryo size.
Following the method in Refs.~\cite{wangklein,SarahIvan,FrenkelAuer,romano}, 
the MC procedure consists of first noting at iteration step $i$ the value of the constraint for a configuration $o$, $\phi_o$, and then generating an unbiased MC trajectory in the 
canonical ensemble with the Metropolis algorithm for 10 displacement attempts per particle to arrive at a new configuration $w$ with a value of the constraint potential
$\phi_w$.  The new configuration is accepted ($w$ becomes the configuration at iteration $i+1$) with probability $\min{\left[1,\exp{\left( \beta \phi_o - \beta \phi_w \right)}\right]}$.  
Otherwise, $o$ remains the configuration at iteration $i+1$.

We carry out biased simulations for several values of $n_0$ for each $T$, and correct for the bias in determining portions of 
$N(n)$ and $P_{\rm max}(n)$ around each $n_0$ according to Ref.~\cite{FrenkelAuer}.  As in Ref.~\cite{romano}
we discard histogram bins with 
poor statistics and simply shift the different portions of $\beta \Delta F(n)$ and $\beta \Delta G(n)$ to minimize the difference in the range of $n$ for which the pieces overlap.
We check our procedure with MBAR~\cite{MBAR} and our results agree to within error.
$\beta \Delta F(n)$ is normalized according to Eq.~\ref{eqpmaxnorm} and for
$\beta \Delta G(n)$, we determine $N_t$ so that $\exp{[-\beta \Delta G(0)]} + \sum_{i=1}^{n^*} \exp{[-\beta \Delta G(i)]} = N_p$.  This latter condition is usually indistinguishable to
within $0.1 k_{\rm B}{T}$
from imposing the condition $\beta \Delta G(0) = 0$ in terms of determining $\beta \Delta G^*$.

When the barrier is sufficiently low, we impose a simple ``hard wall'' constraint, namely, that any MC trajectory that results in $n_{\rm max} >  n_0$ is rejected, using only a single $n_0$ for a given $T$.
When using a hard wall constraint, it is important to not place it much beyond the critical embryo size.  
A good check is that the time series of $n_{\rm max}$ should not get ``stuck'' near the constraint.
If the constraint is too large, poor sampling will result in an apparent barrier height and critical size that are both too large.
In both biasing schemes, we generally use twenty independent starting configurations in order to obtain good averages.

\section{Results}

\subsection{The melting temperature}

\begin{figure}[h]
\centering\includegraphics[clip=true, trim=0 0 0 0, width=8.0cm]{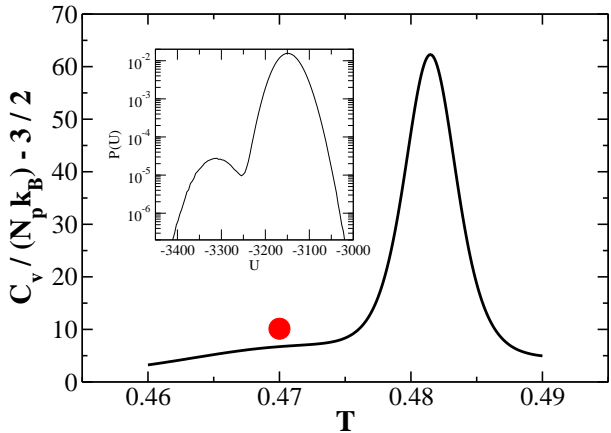}
\caption{
Heat capacity as a function of  $T$.  $C_V(T)$ is determined by reweighting the double-peaked potential energy probability distribution $P(U)$ at $T=0.490$ shown in the inset. The peak at $T=0.482$ marks the coexistence temperature between liquid and solid cluster states. 
The circle is the value of $C_V$ at $T=0.470$ determined from solidified cluster states.}
\label{fig1rev}
\end{figure}

There are two melting temperatures to speak of.  According to Eq.~\ref{eqwork-min}, the barrier to nucleation becomes
infinite and the rate is zero at $T_m$, when $\Delta \mu=0$, i.e., the melting temperature in the 
thermodynamic limit.  For bulk LJ at $p=0$, this is the fcc melting temperature of $0.618$~\cite{baidakov2}.
For comparison, the pressure of our system, evaluated from the virial as for a bulk system, is less than $10^{-4}$, effectively zero.
Thus, $T_m=0.618$ is a reasonable estimate.

However, for our finite-sized cluster, the presence of a surface complicates matters, and the 
coexistence temperature should be defined as the temperature at which the droplet as a whole has equal probability of being either in the solid or liquid state.
To determine this temperature, we note that the system at $T=0.490$ is predominantly in the liquid
state but makes short excursions to being largely solid (a surface melted state). This flipping between states is readily apparent in any of the 501 potential energy time series we have collected (not shown).
From these time series, we construct the probability distribution for the potential energy $P(U)$, which has a distinctly bimodal character
as shown in the inset of Fig.~\ref{fig1rev}.  The main part of the figure shows the heat capacity $C_V(T)$ extrapolated through straightforward temperature reweighting of $P(U)$.  Also plotted is a point for $C_V(T=0.470)$, as determined solely from energy fluctuations in the crystallized state at that $T$. That the discrepancy is small at $T=0.470$ allows us to estimate the coexistence temperature for our cluster
to be $T_m^c=0.482$. Clearly, $T_m^c$ is not the intended melting temperature in Eq.~\ref{eqstar}.

\subsection{Nucleation rates from MFPT}

\begin{figure}[h]
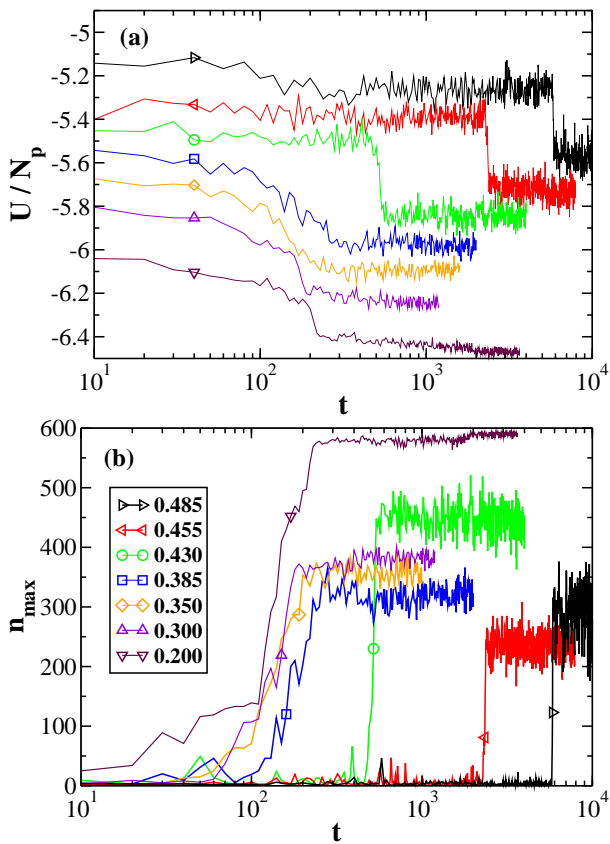

\centering\includegraphics[clip=true, trim=0 0 0 0, width=8.0cm]{fig2a}
\centering\includegraphics[clip=true, trim=0 0 0 0, width=8.0cm]{fig2b}
\caption{Time series of (a) potential energy $U$ and (b) largest embryo size $n_{\rm max}$ showing crystallization
events.  At higher $T$, the nearly vertical changes in  the graphs indicate very fast growth compared to the 
lifetime of the metastable liquid state.  Legend indicates $T$ for both panels.  At  $T=0.385$, the metastable state becomes difficult to discern.
At $T=0.200$, the system progresses essentially monotonically to the frozen state.
}
\label{fig2rev}
\end{figure}

Prior to determining the rate, we consider the potential energy per particle $U/N_p$ as a function of time after the quench from $T=0.530$ to the various target temperatures.  At low to moderate
supercooling, e.g., from $T=0.485$ to $T=0.430$ in Fig.~\ref{fig2rev}(a), the initial rapid change in $U$ shows the system reaching a metastable equilibrium, 
where the droplet is liquid.  The sharp drop in $U$ for these $T$, after metastable equilibrium is achieved, marks rapid growth of a postcritical crystalline embryo, 
as evidenced by the commensurate sharp increase in $n_{\rm max}$ in Fig.~\ref{fig2rev}(b).
At $T=0.385$, the metastable state is less clearly seen, if at all, near $t=60$ and the decrease in $U$ beyond $t\approx90$ is accompanied by an increase in $n_{\rm max}$.
By $T=0.200$, the system proceeds monotonically from the $T=0.530$ state, with both $U$ and $n_{\rm max}$ sliding towards the frozen state.  The sharp change in 
$U$ and $n_{\rm max}$ near $t=200$ occurs after most of the droplet is already crystalline.  While this is interesting, we do not consider it in this study.

\begin{figure}[h]
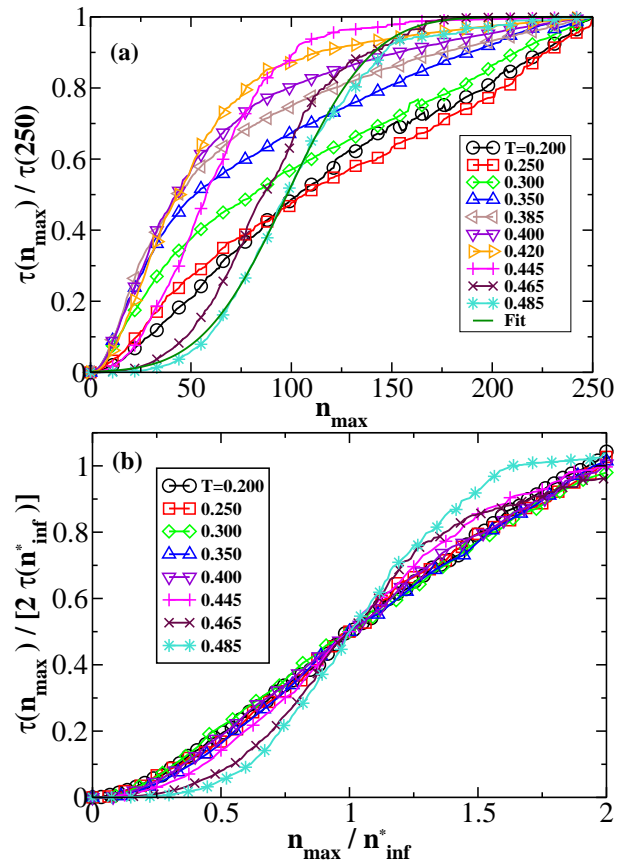

\centering\includegraphics[clip=true, trim=0 0 0 0, width=8.0cm]{fig3a}
\centering\includegraphics[clip=true, trim=0 0 0 0, width=8.0cm]{fig3b}
\caption{Mean first-passage time $\tau(n_{\rm max})$ for the appearance of an embryo of size $n_{\rm max}$ for a range of $T$ indicated by the legend.
In (a) curves are normalized by $\tau(250)$.
For $T=0.485$, we show a fit according to Eq.~\ref{eqMFPT}.
This sigmoidal shape is progressively lost with increased supercooling as the early time plateau shortens.
Below $T=0.35$, curves become less steep at small $n_{\rm max}$, which tends to move the inflection point to larger $n_{\rm max}$, and the curves
become more linear.  In (b) we plot the data rescaled with $n^*_{\rm inf}$, the inflection point. 
}
\label{fig3rev}
\end{figure}

Next we wish to quantify the rate of nucleation from $\tau(n)$.
A sampling of curves from our range of $T$ is shown in Fig.~\ref{fig3rev}, 
where we have normalized the curves by $\tau(n=250)$ since nucleation times vary widely.
We define a crystallization rate as $J_{250}\equiv 1/\tau(250)$ that should approximately 
equal the nucleation rate at shallow supercooling, but clearly underestimate the nucleation rate at  low $T$ as it captures
much time spent by a post-critical embryo growing to a size of 250.

At shallow to moderate supercooling $\tau(n)$ is fairly well approximated by Eq.~\ref{eqMFPT}.
We thus define  $J_{\rm MFPT}\equiv 1/\tau_J $, where $\tau_J$ is determined from fitting to Eq.~\ref{eqMFPT} for $T=0.415$ and higher.  From the fit, we also obtain $n^*_{\rm MFPT}$ as an estimate for $n_F^*$. 

In order to extend the determination of the nucleation rate to lower $T$, we find the inflection point in $\tau(n)$ and so define
$n^*_{\rm inf}$ according to Eq.~\ref{inflection}.  Since the system has equal probability of growing or shrinking at $n_F^*$, we
define another estimate of the nucleation rate $J_{\rm n^*}\equiv 1/[2 \tau(n^*_{\rm inf})]$.  
We plot $n^*_{\rm MFPT}$ and $n^*_{\rm inf}$ in Fig.~\ref{fig9}(b).

The progression of the change of shape of $\tau(n)$ upon lowering $T$ is noteworthy.  At first, the low-$n$ plateau shrinks as $n_F^*$ decreases.  Along with this, the steepness of $\tau(n)$ for small $n$ increases.  However, below $T\approx 0.4$, the curves become progressively less steep, and by $T=0.250$, the inflection clearly occurs at larger $n$.  An increase in $n^*$ on lowering 
$T$ is not predicted by CNT, but rather is predicted by mean field theories of spinodal-type nucleation.  While this warrants further
investigation, we note that there are likely strong non-equilibrium effects at this very low $T$.

\begin{figure}[h]
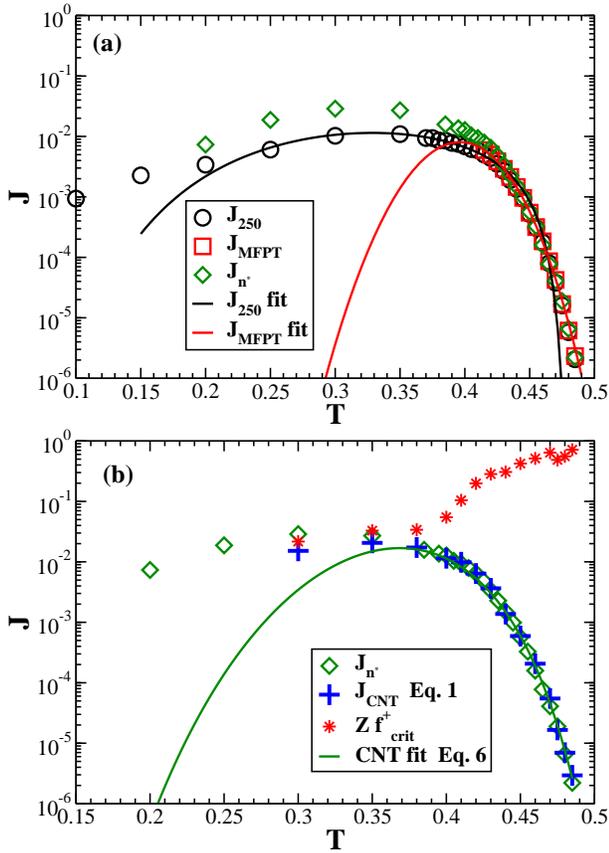

\centering\includegraphics[clip=true, trim=0 0 0 0, width=8.0cm]{fig4a}
\centering\includegraphics[clip=true, trim=0 0 0 0, width=8.0cm]{fig4b}
\caption{
Nucleation rate as a function of $T$.  Panel (a) shows three estimates as described in the text of $J$ based on $\tau(n)$, which all agree at higher $T$.  Curves are 
fits according to Eq.~\ref{eqstar}.  Panel (b) shows a comparison of $J_{\rm n^*}(T)$ with the rate predicted by Eq.~\ref{eqCNTrate} (plus signs) and a
one-parameter fit to obtain $\gamma=0.128$ using Eq.~\ref{eqstar} (curve) with other parameters determined independently.  Also shown is the kinetic prefactor (stars)
of Eq.~\ref{eqCNTrate}.
}
\label{fig4rev}
\end{figure}

We show the temperature dependence of our three rates $J_{250}$, $J_{\rm MFPT}$ and $J_{n^*}$ in Fig.~\ref{fig4rev}.  All three
rates agree from high $T$ down to 0.415, the lowest $T$ at which we determine $J_{\rm MFPT}$.  Below this $T$, 
the difference in $J_{250}$ and $J_{n^*}$ reflects the lack of separation of growth and nucleation time scales.  Both $J_{250}$ and $J_{n^*}$ exhibit a broad
maximum, and show only a weak $T$ dependence below $T=0.4$.

In the next section, we determine the extent to which simple CNT can quantitatively account for the $T$ dependence of the rate.

\subsection{$T$ dependence of the rate from CNT}\label{Trate}

As discussed  in Section II.A, the simplest model for $J(T)$ assumes an Arrhenius dependence of the attachment rate on $T$,
a constant surface tension and a constant difference in enthalpy between the solid and liquid phases.  The resulting model is given
in Eq.~\ref{eqstar}. 
We use it to fit $J_{250}$ and $J_{\rm MFPT}$. 
$J_{250}$ is a crystallization rate blind to the separation of nucleation and growth time scales and should not yield good results.  
By contrast, $J_{\rm MFPT}$ represents a $T$ range for which nucleation and growth are well separated.

Given the orders-of-magnitude difference in the rates as $T$ varies, 
we fit by first taking logarithms of both sides of Eq.~\ref{eqstar}.
The resulting fits of $J_{250}$ and $J_{\rm MFPT}$ are plotted in 
Fig.~\ref{fig4rev}(a), and the fit parameters are as follows.
For $J_{250}$ (fitting from $T= 0.200$ to 0.485):    
$\lambda=87$,
$T_m=0.54$,
$B=1.5 \times 10^{-2}$,
$C=1.7$.
For $J_{\rm MFPT}$ (fitting from $0.415$ to 0.485):    
$\lambda=2.8\times 10^{21}$,
$T_m=0.67$,
$B=0.54$,
$C=14$.
Choosing data from $J_{250}$ in the same temperature range
over which $J_{\rm MFPT}$ is calculated produces similar fit parameters to those for $J_{\rm MFPT}$.
The fits for $J_{\rm MFPT}$ are  more stable with respect to data sampling.
Thus, the parameters vary widely according how much of the data below $T\approx 0.43$ is taken for fitting. 
Unfortunately, fitting yields physically unrealistic or imprecise  parameters.

So while as a fitting function Eq.~\ref{eqstar} is able to reproduce the $T$ dependence of the rate,
it is difficult to extract meaningful physical quantities from the fits parameters.
Our goal is therefore to reduce the fit parameters to just $\gamma$ by independently determining 
$T_m$, $\Delta H$, $A$, $f_0$ and $C$.

\subsubsection{The enthalpy difference}

The enthalpy difference $\Delta H = U_L - U_S + P (V_L - V_S)$ between solid and liquid enters into
the coefficients of Eq.~\ref{eqstar}.
Given that our system is at a very small pressure, that the densities of liquid and crystal are comparable and
that there is a sizeable potential energy difference between liquid and crystal, we approximate $\Delta H \approx U_L - U_S \equiv N_p \Delta u$, where 
$\Delta u$ is the per particle potential energy difference between the liquid and crystal.  The scenario is complicated here by the fact that when our
droplet solidifies, it does so incompletely and remains partially liquid.  Calling $\Delta U$ the difference in potential energy between the liquid and (partially) solidified
droplet, and $\alpha$ the fraction of particles in the solidified droplet identified as solid-like, then we can estimate the enthalpy difference as,
\begin{eqnarray} 
\frac{\Delta H}{N_p} = \Delta u =\frac{1}{\alpha}\frac{\Delta U}{N_p}.
\label{eqdelu}
\end{eqnarray}

\begin{figure}[h]
\centering\includegraphics[clip=true, trim=0 0 0 0, width=8.0cm]{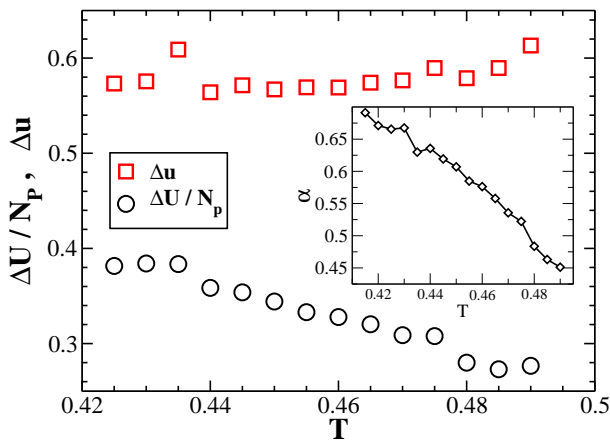}
\caption{Determining $\Delta H/N_p$.  
Circles show the raw estimate $\Delta H = \Delta U$, the system potential energy difference before and after crystallization occurs,
as well as a more refined estimate $N_p \Delta u = \Delta H$ that takes into account $\alpha$ (inset, see Eq.~\ref{eqdelu}) in determining energy differences between solid and liquid particles (squares).  $\Delta u$ is approximately constant with $T$.
}
\label{fig5}
\end{figure}

In the inset of Fig.~\ref{fig5}, we plot $\alpha$ as a function of $T$, and see that the fraction of solid-like particles in the frozen state, at least according to our order parameters, 
increases roughly linearly with decreasing $T$.  In the main panel of Fig.~\ref{fig5}, we plot both $\Delta U/N_p$ and the resulting $\Delta u$.  
We see that
the assumption of constant enthalpy difference between liquid and crystal used in deriving Eq.~\ref{eqstar} is vindicated, and its value is approximately
$\Delta H /N_p = \Delta u = 0.58$.

\subsubsection{Embryo shape}

As noted above, we assume that the surface area of a crystalline embryo within the droplet has surface area $S=A n^{2/3}$. If we assume
spherical embryos and a volume per particle to be that of an fcc particle, $v_{\rm fcc}=1.04$~\cite{depablo}, we obtain $A=4.96$.  To obtain
a better estimate of the shape factor, we model the embryo as an ellipsoid~\cite{trudu, wangklein}.  To do this, we first compute the moment of inertia tensor for all 
particles in the largest embryo in the system. The eigenvalues of this tensor yield the three principal axes lengths 
and hence the surface area of the ellipsoid.

\begin{figure}[h]
\centering\includegraphics[clip=true, trim=0 0 0 0, width=8.0cm]{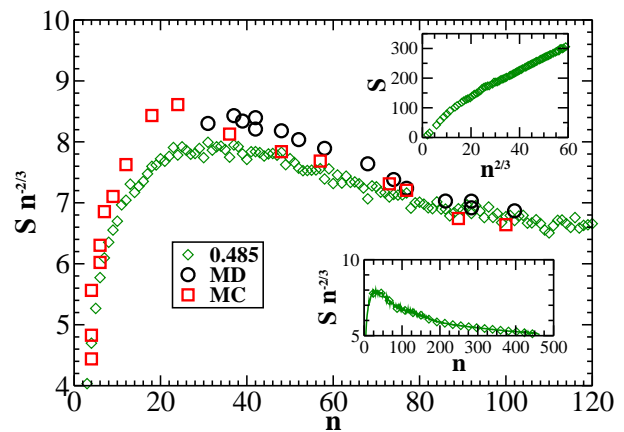}
\caption{
Estimating the shape factor  $A=S n^{-2/3}$ as a function of embryo size, where embryo area $S$ is that of an ellipsoid 
with equivalent moments of inertia as an embryo.  Shown are data for critical clusters from MD ($T\ge 0.410$, circles) and MC (all $T$, squares), 
as well as from all clusters from MD simulations at $T=0.485$.    
In the $T$ range where we expect Eq.~\ref{eqstar} to be valid, corresponding to
$50 < n < 100$, $A$ ranges from about 6.7 to 8.5.  Insets show $S$ as a function of $n^{2/3}$ (upper)
and that $A$ approaches a spherical value of 5 for large $n$ (lower).}
\label{fig6}
\end{figure}

We plot $A=S n^{-2/3}$ as a function of $n$ in Fig.~\ref{fig6}
for both critical embryos from MC (all $T$) and MD ($T\ge 0.410$),
and all $n_{\rm max}$ embryos from MD trajectories for $T=0.485$.  We see that, roughly speaking, the critical embryos
from different $T$ follow the same behaviour as embryos (pre-critical, critical and post-critical) at $T=0.485$.  
For large embryos (shown in the lower inset)
$A$ tends to the spherical value of $\sim 5$, as is expected.  For our range of $T$ of interest (0.415 to 0.485), we see that the embryos become less 
spherical with decreasing size, and that the values of $A$ range from about 6.7 to 8.5 (corresponding to $50 < n < 100$), with an average of 7.6.  The upper inset shows that the dependence of $S$ on $n^{2/3}$ possesses only a slowly varying departure from linearity.

\subsubsection{Attachment rate}

\begin{figure}
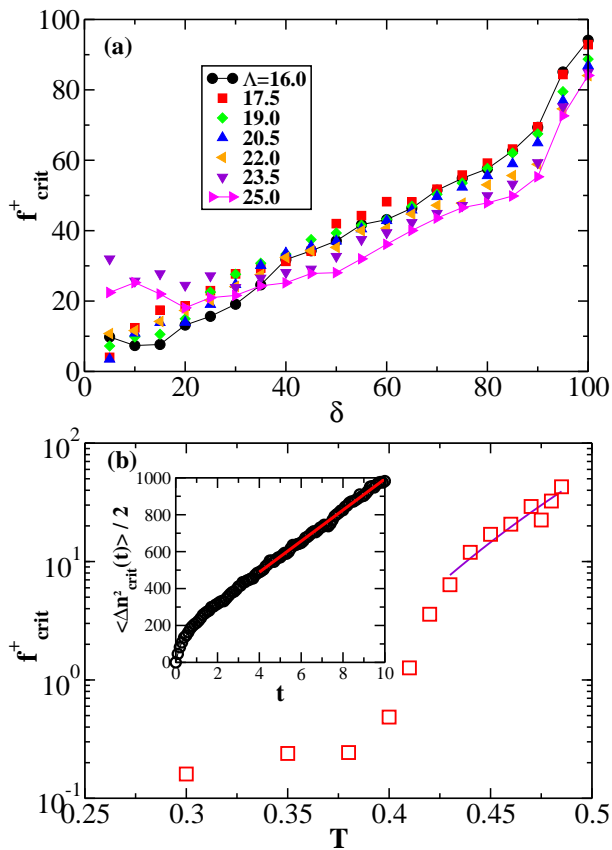

\centering\includegraphics[clip=true, trim=0 0 0 0, width=8.0cm]{fig7a}
\centering\includegraphics[clip=true, trim=0 0 0 0, width=8.0cm]{fig7b}
\caption{
Determination of the attachment rate to the critical cluster.
(a) The effect of $\Lambda$ and $\delta$ on $f_{\rm crit}^+$ for $T=0.485$.  
Short MD trajectories used to determine $f_{\rm crit}^+$ contribute to the average  in
Eq.~\ref{dn2} if $\left|n_{max}(\Lambda)-n_{max}(0)\right| <\delta$.  
Values of $\Lambda$ for the different curves are given in the legend.  To obtain $f_{\rm crit}^+$, we
average over all $\Lambda$ and $30 \le \delta \le 90$. 
(b)  $f_{\rm crit}^+$ (symbols) as a function of $T$.  Solid curve is an Arrhenius fit
 $\ln{f^+_{crit}}=16.4 - 6.2 \frac{1}{T}$ over $0.430\le T\le0.485$.
 Inset shows determination of $f_{\rm crit}^+=84$ for $\delta=100$ and $\Lambda= 25$.
}
\label{fig7}
\end{figure}

To estimate $f_0$, which is essential in the prefactor in Eq.~\ref{eqstar} and defined in Eq.~\ref{eqfcritT}, 
we 
follow Refs.~\cite{FrenkelAuer,Frenkelsalt}.  This method makes use of the fact that the change in size of
a critical embryo follows a simple diffusive process since the free energy landscape is locally flat at the top of the free energy
barrier.  One defines the mean of the squared deviation from the critical size as a function of time,
\begin{equation}\label{dn2}
\left< \Delta n^2(t) \right> = \left< \left[n_{\rm max}(t) - n_{\rm max}(0) \right]^2\right>,
\end{equation}
where $n_{\rm max}(0)=n^*$.  After a very short time, $\left< \Delta n^2(t) \right>$ enters a diffusive regime~\cite{stishovite}, 
i.e., it becomes linear in time, and one obtains in this regime,
\begin{equation}
f_{\rm crit}^+ = \frac{1}{2} \, {\rm slope \,\, of} \,\,  \left< \Delta n^2(t) \right>.
\end{equation}

The usual process is to select a few system configurations containing an embryo of size $n^*$ from MC simulations and to use those
as starting points for MD simulations.  One then selects trajectories that diffuse near $n^*$ and averages over these trajectories, i.e.,
one rejects runs for which the embryo slips off the top of the barrier and shows rapid growth or decay.  
For low barriers, attachment of clusters
of particles to the critical embryo (or break-up of a tenuously-linked embryo), rather than single particle events, may contribute to rapid growth or decay.  
We follow the same procedure, employing from 50 (at low $T$) to 300 (at high $T$) MC configurations.  The criteria for choosing what constitutes
diffusive motion is unclear, for even an embryo that appears to grow rapidly first undergoes a diffusive process, and this diffusive behaviour should be included in the averaging.

To systematically explore this, we define two parameters, $\delta$ and $\Lambda$, and perform averaging in Eq.~\ref{dn2}
for trajectories that satisfy $\left|n_{max}(\Lambda)-n_{max}(0)\right| <\delta$.  In principle, $\delta$ should be of the size over which
the free energy barrier is flat.  $\Lambda$ governs the length of time over which a
trajectory ends up back within $\delta$ of $n^*$.  A small $\Lambda$ eliminates embryos that exhibit large changes in short times, while 
a large $\Lambda$ allows embryos that grow or shrink to return to the critical region.  
Ideally, there should be a range of $\delta$ and $\Lambda$ over which $f_{\rm crit}^+$ is invariant.  We note that we employ averaging
over time origins, i.e., if an embryo returns to $n^*$ after a time of 4, we treat that time as the beginning of an independent trajectory.

The results for $f_{\rm crit}^+$ as a function of $\delta$ for different $\Lambda$ values for $T=0.485$ 
are shown in Fig.~\ref{fig7}(a).  We see that 
for $\delta<30$, there is a large spread in $f_{\rm crit}^+$ over different $\Lambda$.  For $\delta > 90$, there is a rapid increase in $f_{\rm crit}^+$.
For $\delta$ in between, we see no obvious way to choose an optimal $f_{\rm crit}^+$, and so we average over the range $30 \le \delta \le 90$ over all $\Lambda$ for this $T$ to obtain $f_{\rm crit}^+=43$ with a standard deviation of 13.  
While a value of $\delta=90$ seems to be large, approaching $n^*$ in fact, we note that the time over which
the slope of $\left< \Delta n^2(t) \right>$ is taken is fixed to be from 4 to 10, signifcantly smaller than our smallest $\Lambda$.  Shown
in the inset of Fig.~\ref{fig7}(b) is $\left< \Delta n^2(t) \right>$ for (extreme values) $\delta=100$ and $\Lambda=25$, and it appears to be rather well behaved, therefore 
 not providing grounds for rejection on its own.  
We repeat the examination of $f_{\rm crit}^+$ as a function of $\delta$ and $\Lambda$ for each $T$.  
Our analysis indicates a need for a more refined way of determining $f_{\rm crit}^+$ if more precise values are required.


In this way, we obtain $f_{\rm crit}^+$ across our $T$ range, which we plot in Fig.~\ref{fig7}(b).
It shows a super-Arrhenius decrease with $T$ until an apparent falling out of equilibrium
below $T=0.4$, behavior consistent with typical glassy dynamics of simple liquids.
However, as we are primarily concerned with finding $\gamma$ through Eq.~\ref{eqstar}, the
figure also shows a fit of $f_{\rm crit}^+$ to the Arrhenius behaviour in Eq.~\ref{eqfcritT} over 
$0.430\le T\le0.485$, with fit parameters 
$C=6.2\pm0.3$ and
$f_0=\exp{(16.4\pm0.7)}=1.3\times 10^7$ ($6.6\times 10^6$ to $2.7 \times 10^7)$.  
The marked departure below $T\approx0.40$ from the behavior at higher $T$ is consistent with the liquid not achieving metastable equilibrium.


\subsubsection{Surface tension}

Studies of crystal nucleation in bulk LJ liquid report values of $\gamma= 0.28$ to $0.30$ for $T=0.43$ and 0.45, respectively~\cite{baidakov2},
and these compare favourably with the surface tension of a flat interface at the same $T$~\cite{baidakov}.
Using  our estimates for the parameters other than $\gamma$, namely,
 $\Delta H = 0.58 N_p$,
 $A = 7.6$,
 $f_0 = 1.3\times 10^7$,
 $C = 6.2$,
 and the literature values of $T_m=0.618$~\cite{baidakov2}
 and $\gamma=0.3$~\cite{baidakov2}, we obtain
$B= 2.0$ and
$\lambda=8.6 \times 10^8$.  The resulting curve, according to Eq.~\ref{eqstar} is not plotted because it fails to recover the rates 
in Fig.~\ref{fig4rev}(a) by several orders of magnitude.

Therefore, we proceed to find $\gamma$ from a one parameter fit of $J_{n^*}(T)$ with Eq.~\ref{eqstar}, 
using the above values for the other parameters.
Fitting from $T=0.35$ to 0.485, we obtain $\gamma=0.13$, which is significantly lower than the bulk value.
The value of $\gamma$ is quite robust to how much of the data below $T=0.40$ is used.
The fit is plotted in Fig.~\ref{fig4rev}(b), and models the data well down to $T=0.385$. 
This validates the approximations incorporated into CNT, namely of constant  $\Delta H$, $A$, $f_0$, $C$ and $\gamma$.
We note that although the Arrhenius modelling of $f_{\rm crit}^+(T)$ is valid for $T \ge 0.43$, the fit of Eq.~\ref{eqstar}
is rather good down to $T=0.385$ for two reasons: one, the changes in $J(T)$ are driven largely by changes in 
$\beta \Delta G^*$ and two, the difference between the Arrhenius model and the actual values of $f_{\rm crit}^+(T)$
between $T=0.385$ and $T=0.43$ is maximally of the order of a factor of five, which is compensated by an slight 
overestimation of $\beta \Delta G^*$ by the model.
The departure of the fit from data at low $T$ is due to the dramatic change in behavior of  $f_{\rm crit}^+$ below $T=0.40$.

In the next section, we test to what extent these approximations hold in the context of $\beta \Delta G(n)$.  We also test the 
ability of Eq.~\ref{eqCNTrate} to predict the rate, when $\beta \Delta G(n)$ and the other quantities in the equation are calculated
through MC simulations.

\subsection{Free energy barriers}

\subsubsection{$\beta\Delta G(n)$ from  MC calculations}

\begin{figure}[h]
\centering\includegraphics[clip=true, trim=0 0 0 0, width=8.0cm]{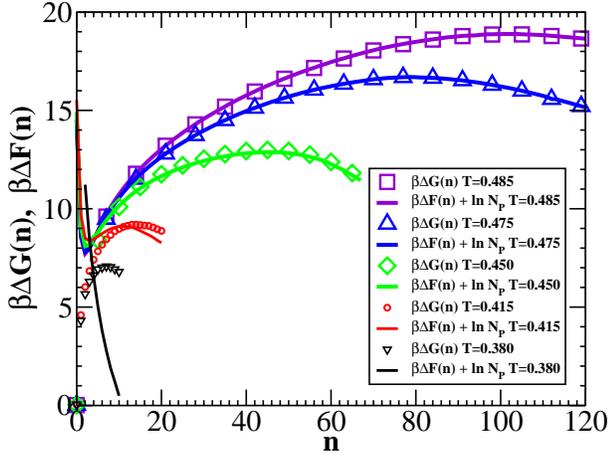}
\caption{
Barrier profiles from umbrella sampling MC for $\beta \Delta G(n)$ (symbols) and $\beta \Delta F(n)$ (curves), which have been shifted
up by $\ln{N_p}$ and which possess a minimum at small $n$.  Below $T=0.405$ (not shown), at which $\beta \Delta G^* \approx \ln{N_p}$, 
$\beta \Delta F(n)$ decreases monotonically.
}
\label{fig8}
\end{figure}

In Fig.~\ref{fig8}, we present a sampling of  the barrier profiles  obtained from MC simulations.  
The $\beta \Delta F(n)$ curves are shifted up by  $\ln{N_p}$ as discussed in Section II.B, and overlap well with the $\beta \Delta G(n)$ for the higher $T$.  Parabolic fits within $\sim k_{\rm B} T$ of the maxima in the curves, allow us to determine $Z$, $Z_F$, $\beta \Delta G^*$, $\beta \Delta F^*$, $n^*$ and $n_F^*$.

Below $T=0.405$, as shown in the figure for $T=0.380$, the $\beta \Delta F(n)$ curves are monotonically decreasing.  The interpretation
of this results is laid out in Ref.~\cite{RegueraCROSS} in the context of the vapour to liquid transition but still above spinodal conditions.
The monotonic decrease means that
for any value of $n_{\rm max}$, it is more probable for $n_{\rm max}$ to increase in size than to decrease.  Thus the system has lost metastability
and unavoidably transforms to the solid.  However, the work of forming a critical embryo is still positive [$\beta \Delta G(n^*) \approx 7.6$].
So while the liquid phase is locally stable against fluctuations towards the solid state, the system as a whole is not, since it is large enough
to make it probable for a critical embryo to appear somewhere in the system on the time scale required for the diffusive attachment of particles.

\subsubsection{$T$-dependence of barrier heights and critical embryo sizes, and rate prediction.}

\begin{figure}[h]
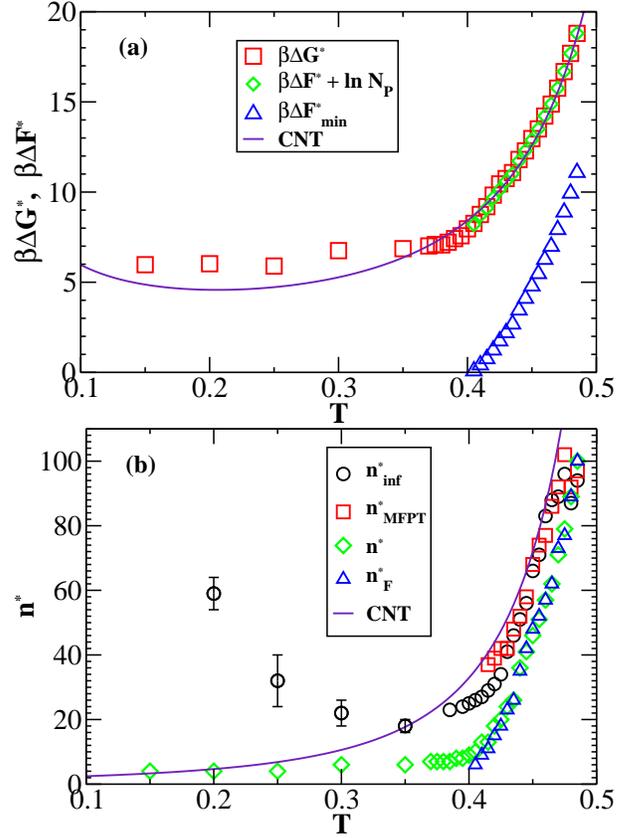

\centering\includegraphics[clip=true, trim=0 0 0 0, width=8.0cm]{fig9a}
\centering\includegraphics[clip=true, trim=0 0 0 0, width=8.0cm]{fig9b}
\caption{
(a) Nucleation barrier heights as a function of $T$. Solid curve shows the prediction based on
CNT after obtaining only $\gamma=0.13$ from a fit to $J(T)$ via Eq.~\ref{eqstar}.
$\beta \Delta F_{\rm min}^*=0$ signals the onset of growth-limited nucleation.
(b) Size of the critical cluster as a function of $T$ from various estimates.  Solid curve is the 
CNT prediction using the same parameters as in panel (a).
}
\label{fig9}
\end{figure}

The $T$ dependence of barrier heights is shown in Fig.~\ref{fig9}(a), while that of critical embryo size in
Fig.~\ref{fig9}(b).  For the barriers, both $\Delta F^* + \ln{N_p}$ and $\Delta G^*$ agree quite closely. 
The crystallization process becomes formally driven by growth-limited nucleation when 
$\beta \Delta F_{\rm min}^*=0$ at $T=0.405$, at which point
$\beta \Delta G^* = 7.6$.
In Ref.~\cite{RegueraCROSS}, the authors gave a simple criterion for the onset of 
growth-limit nucleation, namely that $P_{\rm max}(n^*)\approx 1$, or $\beta \Delta F(n^*)=0$, which implies
$\beta \Delta G^* \approx \ln{N_p}= 6.40$, which is roughly $1 \, k_{\rm B}T$ lower than what we obtain.  But as this is
a rule of thumb, the prediction is quite good.

Below the crossover temperature of $T=0.405$, both $\beta \Delta G^*$ and $n^*$ vary significantly less with 
decreasing $T$.   This trend is consistent with the predictions of CNT shown in Fig.~\ref{fig9}, especially if 
$n^*$ is to remain finite as it appears to do.
The crossover more or less coincides with a flattening out of the $T$ dependence of $f_{\rm crit}^+$, as shown in Fig.~\ref{fig7}(b).
We note that below $T=0.405$, the equilibrium dynamics, if one could probe them, may be quite slow, and the time scale of 
liquid relaxation appears to be significantly longer than the time scale
of embryo assembly, and thus we see an interplay between glassy dynamics and nucleation~\cite{glassynuc}.  
Nucleation below this temperature is occurring in an aging, non-equilibrium liquid, and this
warrants further exploration.

In Fig.~\ref{fig9}(b), we see significant differences in critical embryo size, both between $n^*$ and $n_F^*$ and more strikingly, between
$n_F^*$ (MC) and $n^*_{\rm inf}$ or $n^*_{\rm MFPT}$ (both MD).  This is not a consequence of the definition of what constitutes a solid-like particle, but rather a real difference in the structures accessible to MD and MC in the critical region.  
At low $T$, where we are increasingly out of equilibrium, $n_{\rm inf}^*$ in fact increases as $T$ decreases. 
Even at moderate supercooling, the critical size is larger for MD simulations.

\begin{figure}[h]
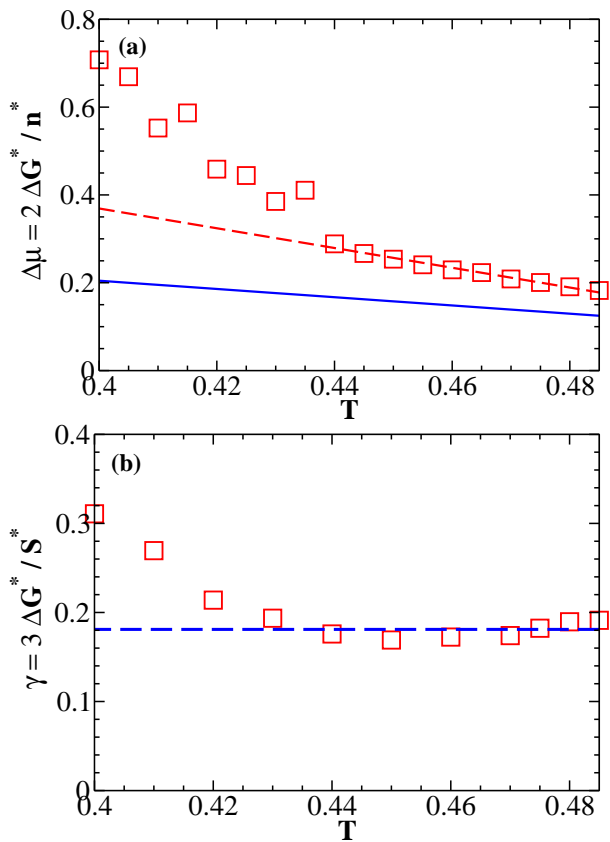

\centering\includegraphics[clip=true, trim=0 0 0 0, width=8.0cm]{fig10a}
\centering\includegraphics[clip=true, trim=0 0 0 0, width=8.0cm]{fig10b}
\caption{
Estimates of $\gamma$ and $\Delta \mu$ from $T$ dependence of MC data.
Panel (a)  shows the CNT relation $\Delta\mu = 2 \Delta G^*/n^*$ as a function of $T$.  
Blue solid line is $\Delta u - \Delta u T/T_m$, setting $T_m=0.618$
and $\Delta u=0.58$.  The dashed line is a fit (for $T\ge0.44$), yielding 
$T_m=0.564$ and $\Delta u=1.27$.
Panel (b), $\gamma = 3\Delta G^* / S^*$ versus $T$.  Dashed line is
a fit  (for $T\ge0.43$) 
with a constant, yielding $\gamma=0.18$.
}
\label{fig10}
\end{figure}

Having calculated $\beta\Delta G(n)$, and hence obtained $Z$ and $n^*$ as well, we can now 
predict $J_{\rm CNT}(T)$ according to Eq.~\ref{eqCNTrate}, and we show the result in Fig.~\ref{fig4rev}(b).
The agreement with $J_{n^*}$ is rather good, showing discrepancy only at $T=0.35$ and below.
Also shown in Fig.~\ref{fig4rev}(b) is the
kinetic prefactor $Z f_{\rm crit}^+$.  Similarly to what was observed  in Ref.~\cite{RegueraCROSS} for 
the vapour to liquid transition, once the growth-limited nucleation regime is entered, the kinetic prefactor
dictates the $T$ dependence of the rate.

Eq.~\ref{eqCNTrate} is the CNT prediction of the rate that lacks any thermodynamic modelling of the 
work of forming a critical embryo.  We have already seen that modelling $\beta\Delta G(n)$ through 
Eq.~\ref{eqwork-min} and estimating the thermodynamic quantities that enter it and Eq.~\ref{eqstar}
matches the rate from MD, but with a smaller value of $\gamma$ than expected.  
Whatever values of $\gamma$ and $\Delta u$ we derive from $\beta\Delta G(n)$ as obtained
from MC, from what we already know, we expect that they should combine to produce nearly equal values of
$\beta\Delta G^*$ as implied from MD
(since both Eq.~\ref{eqCNTrate} and Eq.~\ref{eqstar} recover the rate) and a smaller $n^*$.  This later condition
implies that we should obtain a larger value of $\gamma$.
We also wish to test whether the constancy of $\Delta u$ and $\gamma$ obtained from MD  for $T\ge0.4$ (i.e., from 
 the $T$ dependence of the rate and direct calculation) is borne out in the $\Delta G(n)$ MC data.

To this end, we plot in Fig.~\ref{fig10}(a) for $T \ge 0.4$, 
the quantity $2 \Delta G^*/n^*$, which according to Eq.~\ref{eqwork-min} should equal $\Delta \mu(T)$,
which in turn should be $\Delta \mu(T) = \Delta H ( 1 - T/T_m)/N_p \approx \Delta u - \Delta u \,T/T_m$.
A linear fit to data only for $T\ge 0.44$ looks convincing, and yields $T_m=0.564$ and a value of $\Delta u =1.27$ 
that is significantly higher than the independently calculated value of $0.58$, roughly by a factor of 2.2.
Similar discrepancies have been noted for MC studies of nucleation in Ref.~\cite{romano}, where
across many state points the value of $\beta \Delta \mu$ obtained from fits to Eq.~\ref{eqwork-min} were
a factor of 2.5 higher than those calculated from thermodynamic integration, i.e., the true value.

In Fig.~\ref{fig10}(b), we plot $\gamma = 3 \Delta G^*/S^*$, which again follows from Eq.~\ref{eqwork-min},
where $S^*$ is the area of the critical embryo.
For a good range of data, $\gamma$ is indeed constant.  
A fit to a constant  for $T\ge 0.43$ yields $\gamma =  0.18$, which is higher than what the rate data imply, but still significantly lower than  the expected value of $0.3$.
These MC-derived values of $\gamma$, $\Delta u$ and $T_m$ do not produce a particularly good fit to the rate when plugged in to
Eq.~\ref{eqstar}.

To compare MC-derived parameters and those obtained from MD in another way, we plot, according to CNT (Eq.~\ref{eqwork-min}) predictions, $\beta \Delta G^*$ from Eq.~\ref{eqGstarTCNT} in Fig.~\ref{fig9}(a) and 
$n^* = 2 B k_{\rm B}T_m/[\Delta u(T_m - T)^3]$ in Fig.~\ref{fig9}(b), using parameters as obtained in Section~\ref{Trate} 
($\gamma=0.13$,  $\Delta H = 0.58 N_p$ and $T_m=0.618$, giving $B=0.16$).  We find remarkably good agreement
for $\beta \Delta G^*(T)$ (even for $T < 0.4$) with MC while the CNT expression for $n^*$ gives values that are 
significantly higher than the MC result.  

We conclude from these comparisons that the discrepancies $\gamma$ and $\Delta u$ between MD and MC are 
consistent with a larger $n^*$ from MD, since $\gamma$ from MD is smaller.  However, while from MD we find that 
$\Delta u$ is constant, MC does not show this to the same extent.  Therefore, we also conclude that in order to obtain 
quantitative estimates from the MC-derived $\Delta G(n)$, a more nuanced modelling of $\beta\Delta G(n)$ than in Eq.~\ref{eqwork-min}, and a more careful definition of the surface area of embryos (including more precise definitions of liquid-like and solid-like particles) are required.

\subsection{Escape from the critical state}

\begin{figure}[h]
\centering\includegraphics[clip=true, trim=0 0 0 0, width=8.0cm]{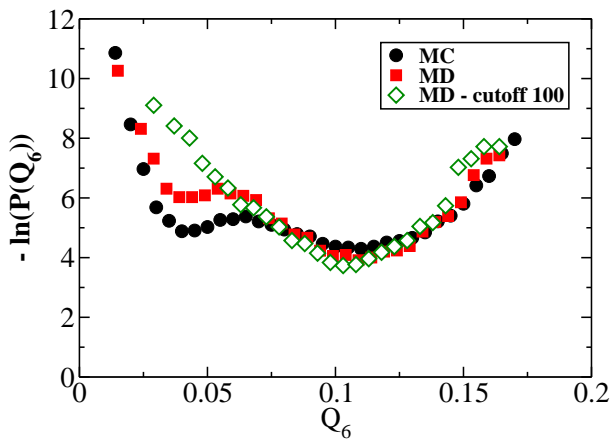}
\caption{Probability distributions for $Q_6$ at $T=0.475$ for $60\le n_{\rm max} \le 100$.  See text for explanation.
}
\label{fig11}
\end{figure}

We now explore the differences in $n^*$ between MC and MFPT results that begin to be felt at $T=0.475$.
According to MFPT, $n^*\approx100$.  In Fig.~\ref{fig11} we plot the probability density $P(Q_6)$ for $Q_6$,
a global measure of the crystallinity of the system as a whole.  We plot the negative of the logarithm of the distribution
in order to view it as a free energy.
Generally speaking, two factors contribute to the value of $Q_6$, the number of crystal-like particles and
the relative orientation of crystal-like domains.  For example, $Q_6$ will grow as the size of an fcc crystallite increases,
but a large icosahedral embryo of similar size consisting of 20 fcc tetrahedra sharing a vertex, will have a lower value 
of $Q_6$.

\begin{figure}
\centering\includegraphics[clip=true, trim=0 0 0 0, width=6.0cm]{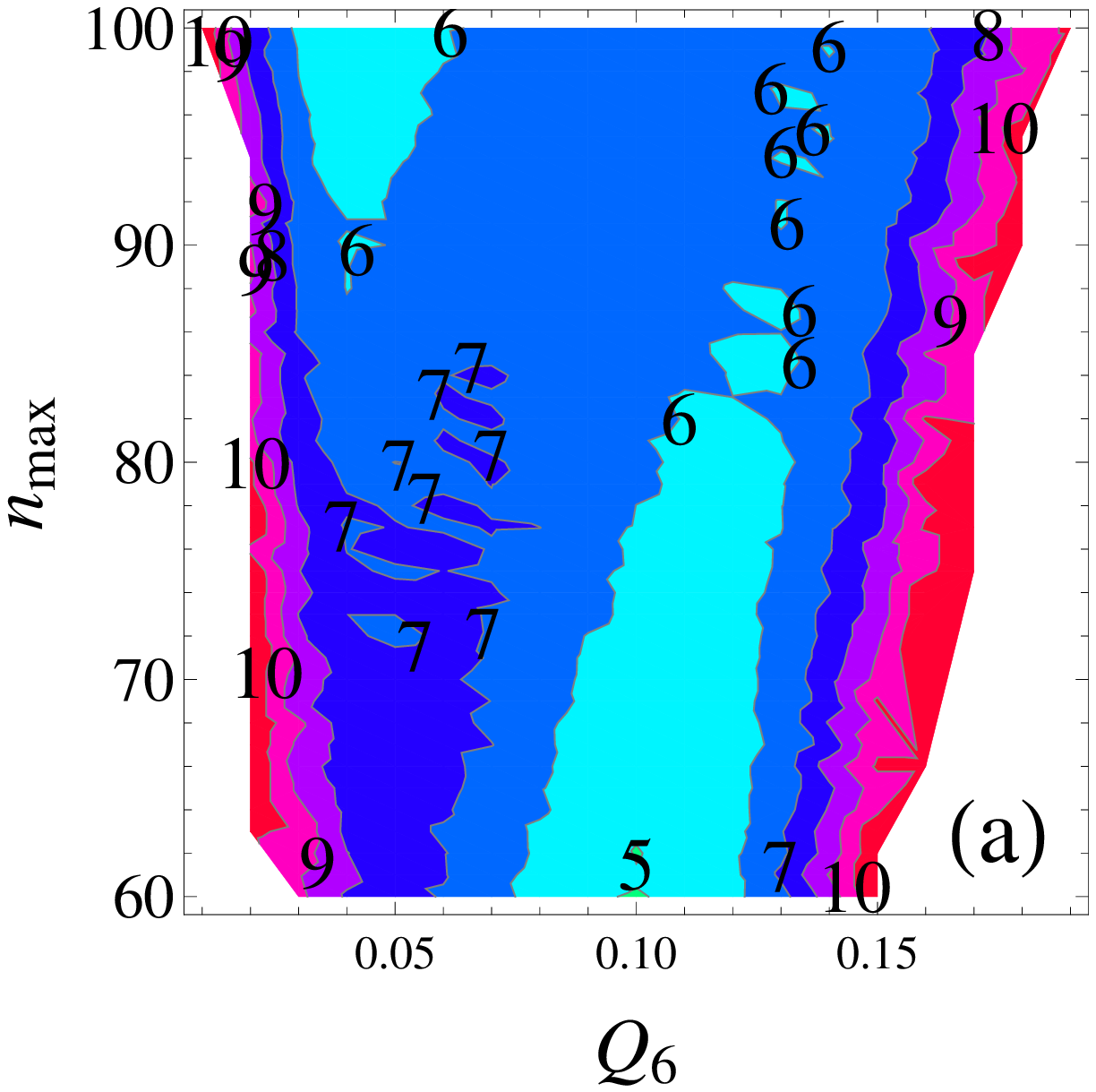}
\centering\includegraphics[clip=true, trim=0 0 0 0, width=6.0cm]{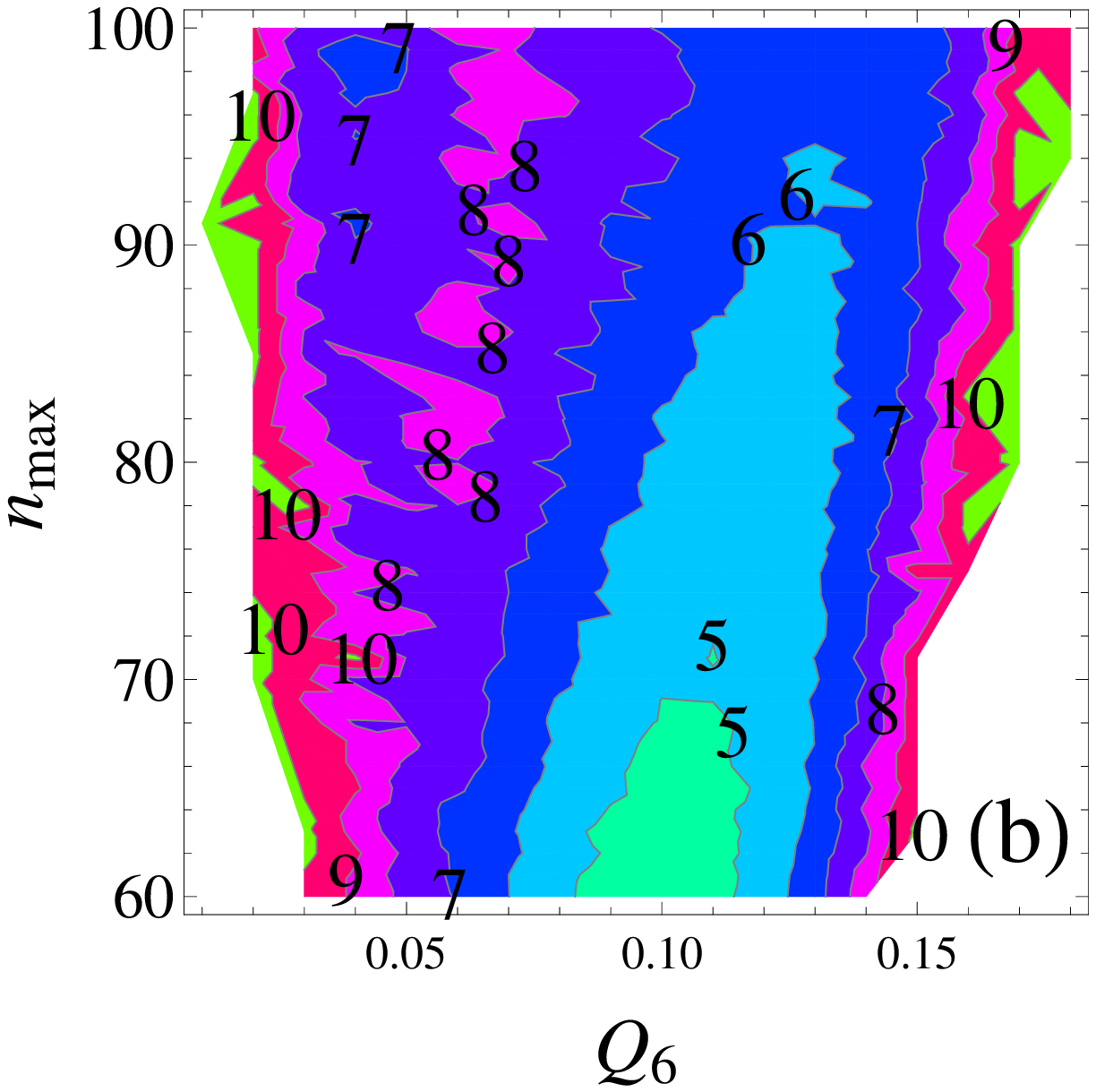}
\centering\includegraphics[clip=true, trim=0 0 0 0, width=6.0cm]{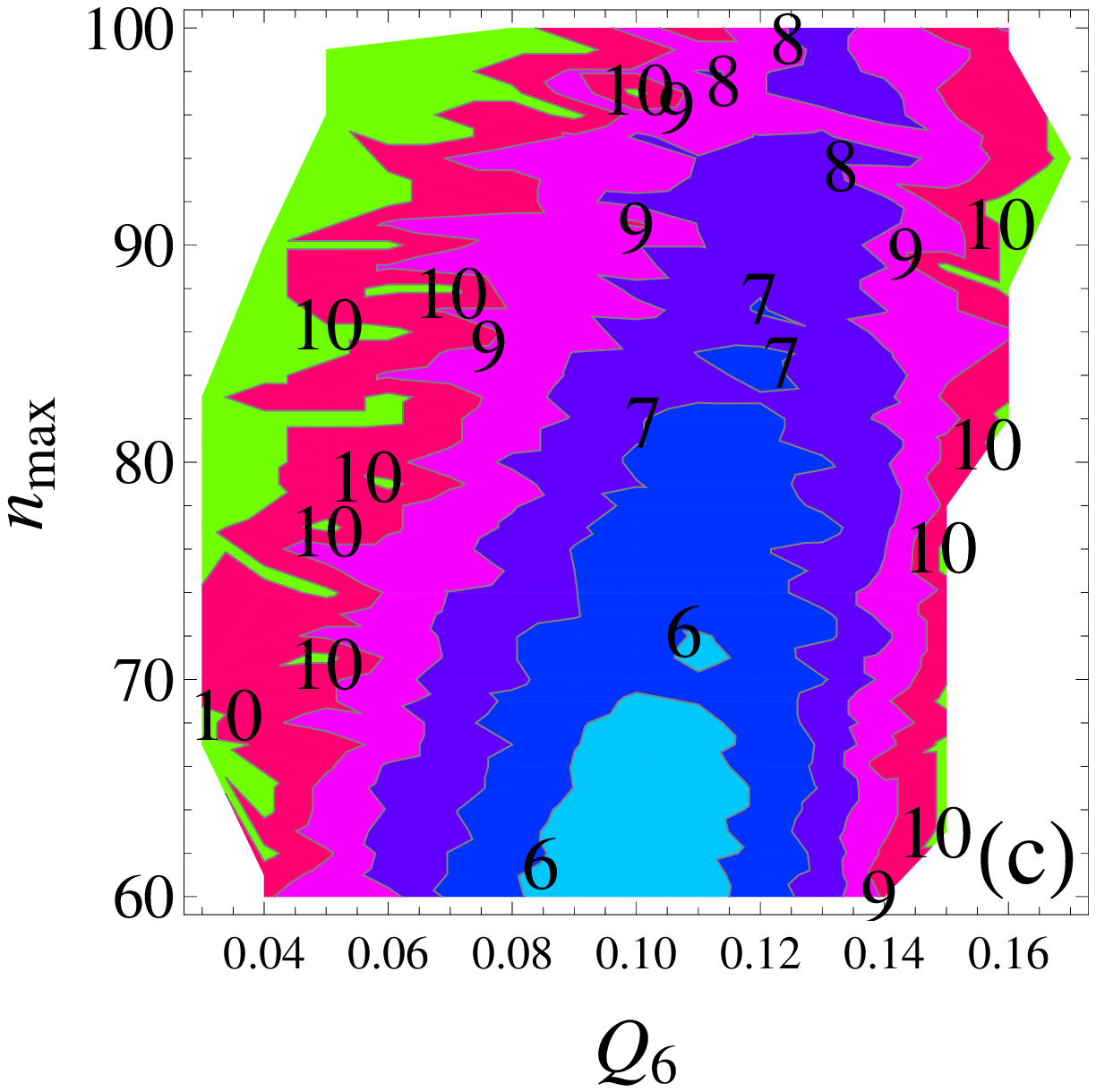}
\caption{Joint probability distributions for  $n_{\rm max}$ and $Q_6$ at $T=0.475$.  
Plotted is $-\ln{P(n_{\rm max},Q_6})$ for (a) MC, (b) MD, (c) MD without allowing retracing to $n_{\rm max}\le 100$. 
The contour lines are in increments of 1.
}
\label{fig12}
\end{figure}

In the first instance we calculate  $P(Q_6)$ from MD crystallization trajectories, using data up to the first time
that $n_{\rm max}$ reaches 100, utilizing all configurations with $60 \le n_{\rm max} \le 100$.  
In this way, we consider embryos in the critical region but do not allow embryos to sample states beyond the critical size.
The result is a unimodal $P(Q_6)$ with a preferred value of $Q_6=0.1$.  We refer to
this value of $Q_6$ as {\it high}.  If we consider embryos from all times along the trajectory, i.e., we allow the system
to sample post-critical states and subsequently shrink back into the pre-critical region, the distribution changes by 
exhibiting a localized preference for  $Q_6 = 0.04$ [a shallow minimum in $-\ln{P(Q_6)}$].  We refer to
this value of $Q_6$ as {\it low}.  Finally, we carry out MC simulations with hard wall constraints to enforce 
$60 \le n_{\rm max} \le 100$.  The resulting fee energy, also shown in Fig.~\ref{fig11}, shows that the
relative preferences for high and low $Q_6$ structures are similar, and that there is a free energy barrier separating the two.  
Thus, although there exist qualitatively different 
equilibrium structures in the critical region (same $n_{\rm max}$, different $Q_6$), MD trajectories do not easily sample the
low $Q_6$ states until after embryos have crossed into the post-critical region.  The kinetics of crossing the small barrier 
for $n_{\rm max} \le 100$  are apparently significantly slower than structural changes occurring for $n_{\rm max} > 100$.

To develop a better picture of the process, we use the data from Fig.~\ref{fig11} to construct two-dimensional probability distributions in both $Q_6$ and $n_{\rm max}$.  The results are plotted in Fig.~\ref{fig12} as contour plots of
$-\ln{P(n_{\rm max},Q_6)}$.  For the equilibrium MC data in panel (a), we see a single trough coming into the critical region 
from $n_{\rm max}=60$ and $Q_6=0.1$ that becomes fairly flat at larger $n_{\rm max}$.
For $n_{\rm max}\ge 90$, there are two exiting troughs: a weak one at high $Q_6$ that continues the incoming one; and
a more dominant one at low $Q_6$.  There is a small ridge separating the two troughs.

Panel (b) of Fig.~\ref{fig12} shows MD data where post-critical embryos that retrace back below $n_{\rm max}=100$
are counted.  The exiting trough at low $Q_6$ is higher in free energy and is much weaker than the high $Q_6$ exiting trough.  The MD data for which no retracing is allowed, in panel (c), show only the high $Q_6$ exiting trough.


Thus, while it is possible for $n_{\rm max}<100$ embryos to transform from high to low $Q_6$, and both states have similar free
energies, as the MC data show, the ridge separating high and low $Q_6$ prevent the MD trajectories from exploring these low $Q_6$ states.  Further, it is clear that the critical embryo size is significantly smaller when low $Q_6$ states are sampled, and this
is responsible for the discrepancy between MC and MD estimates of $n_F^*$.  Another major point is that 
we do not see two competing pathways entering the critical region.  The  low $Q_6$ exiting trough
only forms near the critical region.

\begin{figure}[H]
\includegraphics[clip=true, trim=0 0 0 0, width=3.8cm]{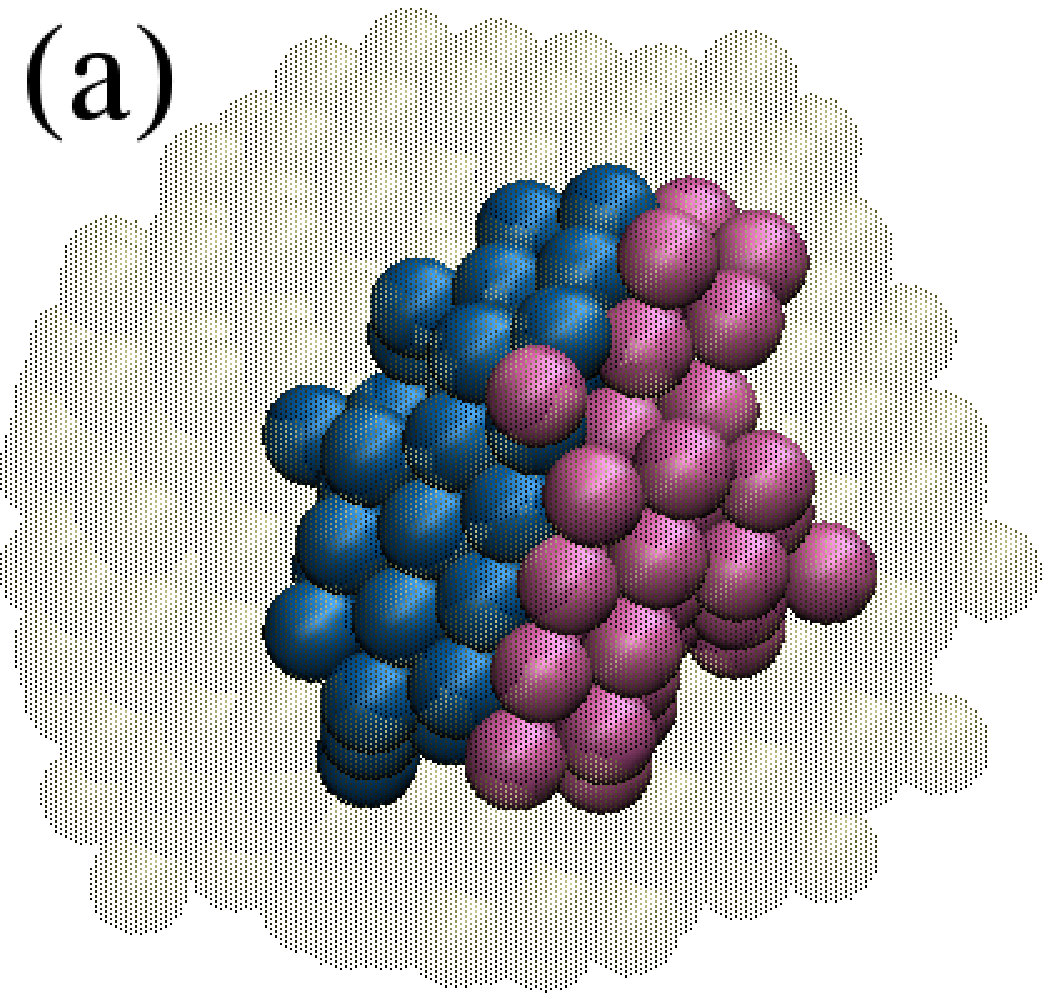}
\includegraphics[clip=true, trim=0 0 0 0, width=3.8cm]{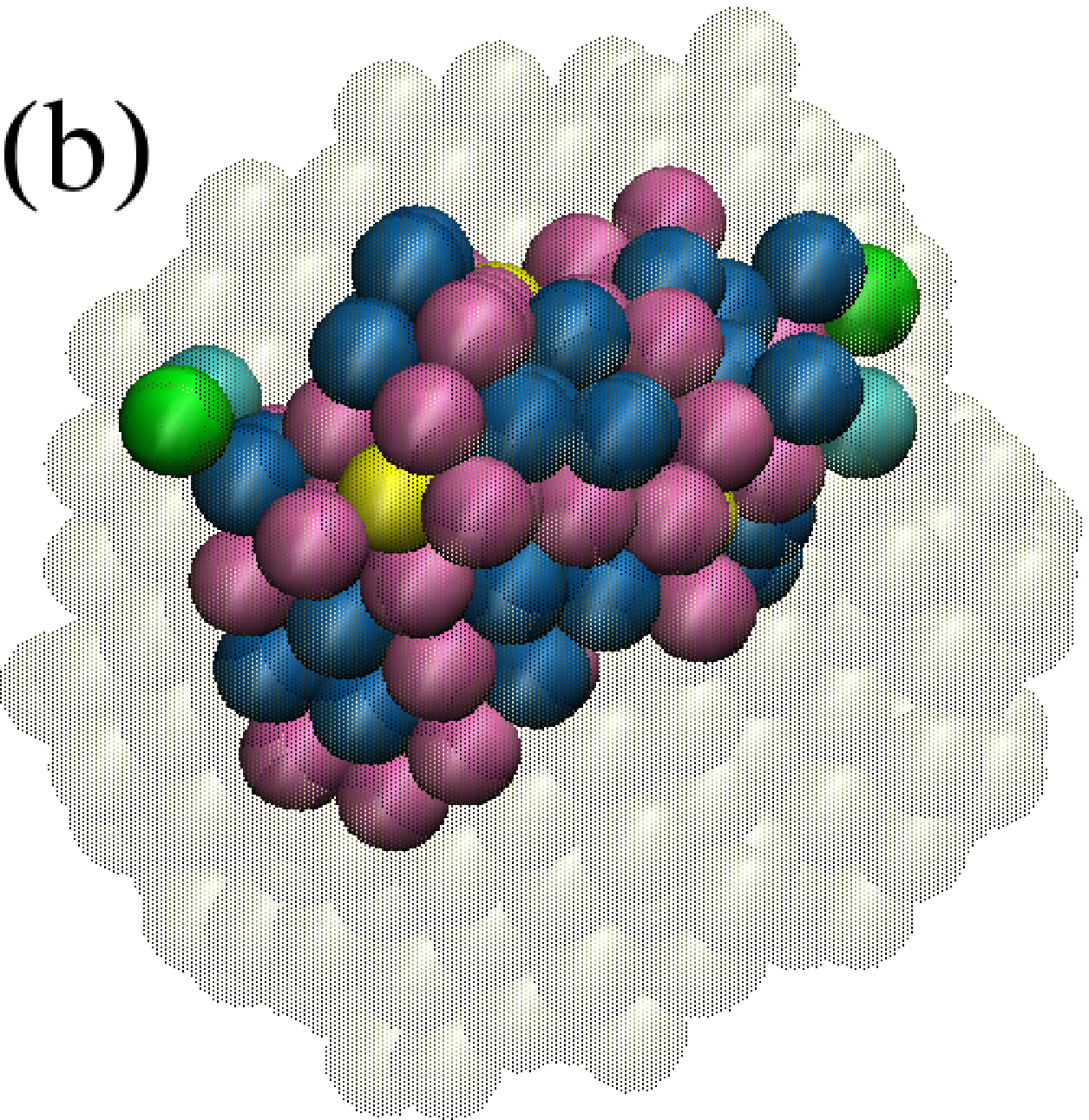}
\includegraphics[clip=true, trim=0 0 0 0, width=3.8cm]{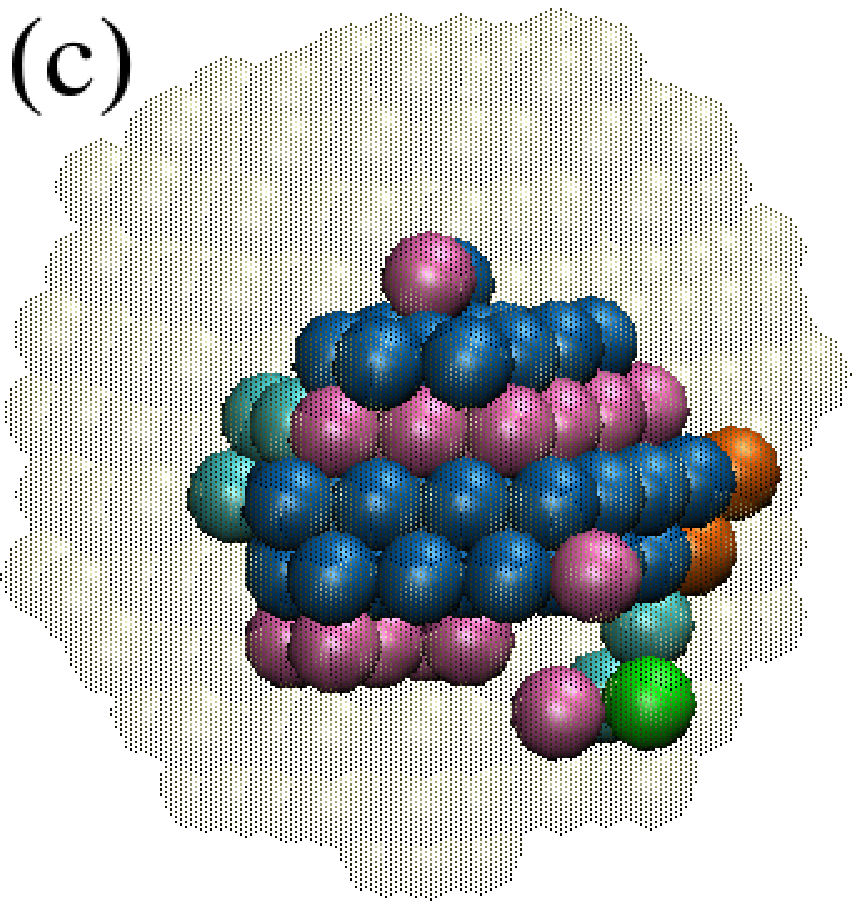}
\includegraphics[clip=true, trim=0 0 0 0, width=3.8cm]{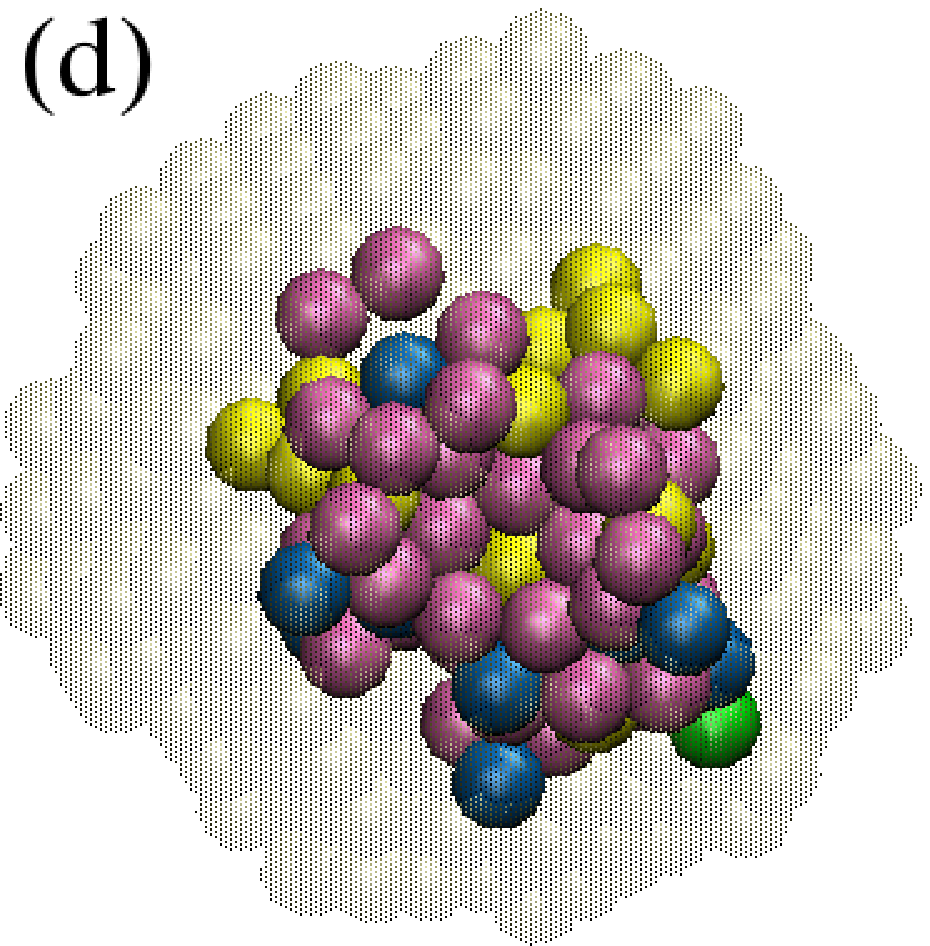}
\includegraphics[clip=true, trim=0 0 0 0, width=3.8cm]{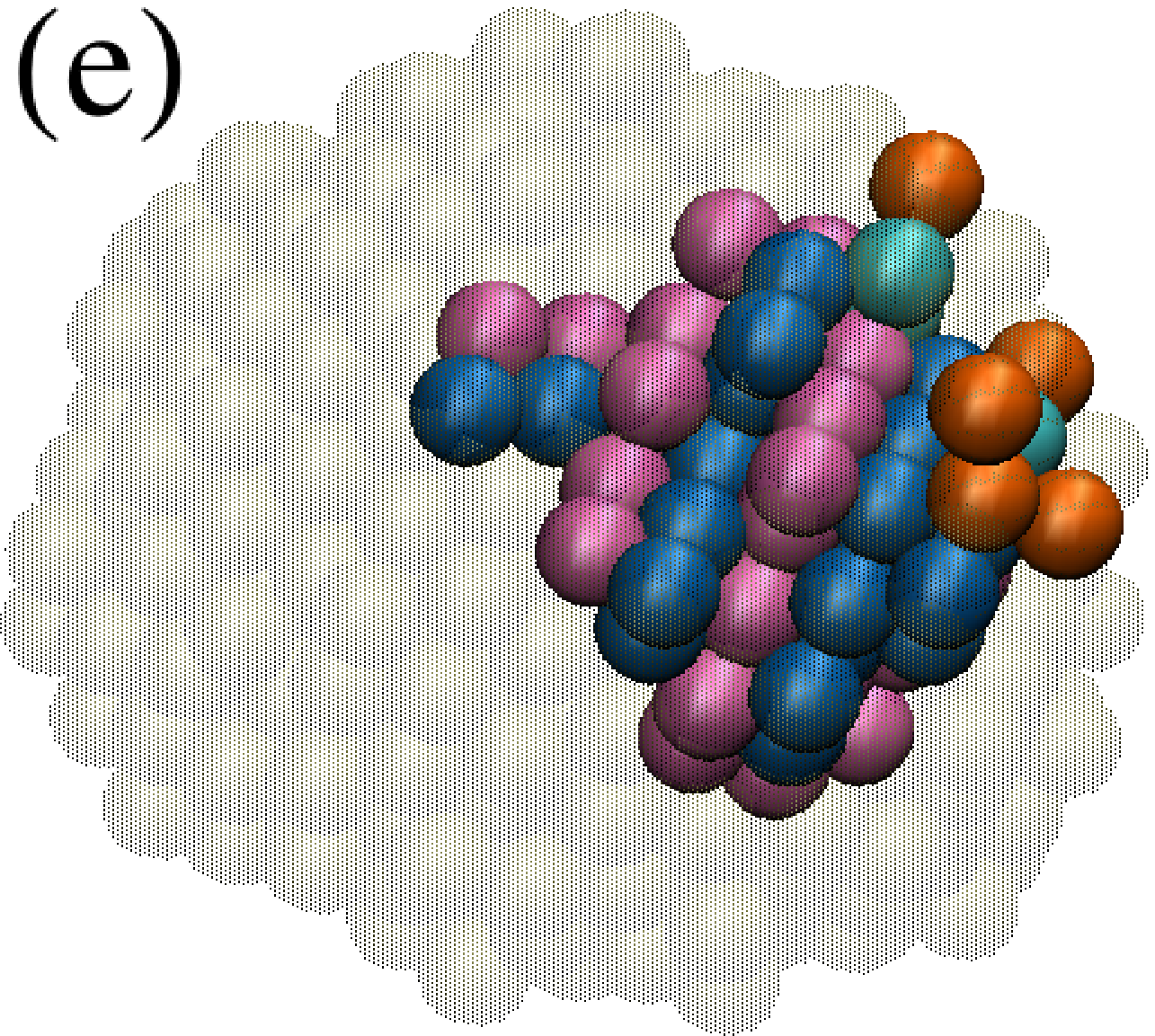}
\includegraphics[clip=true, trim=0 0 0 0, width=3.8cm]{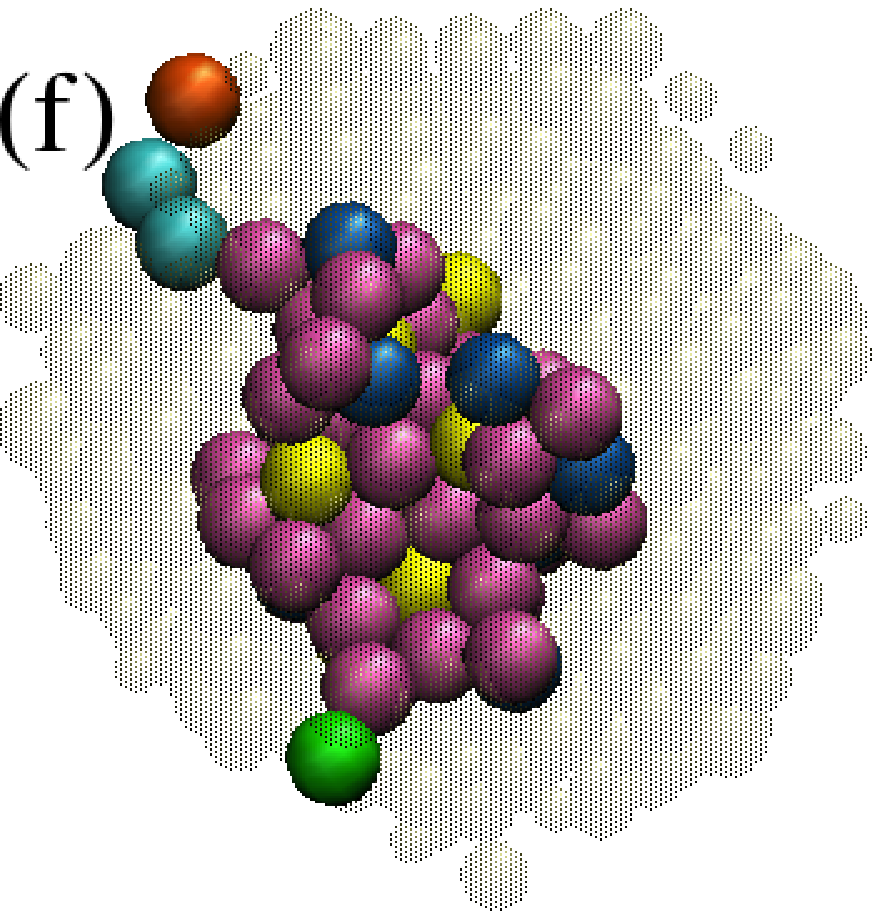}
\includegraphics[clip=true, trim=0 0 0 0, width=3.8cm]{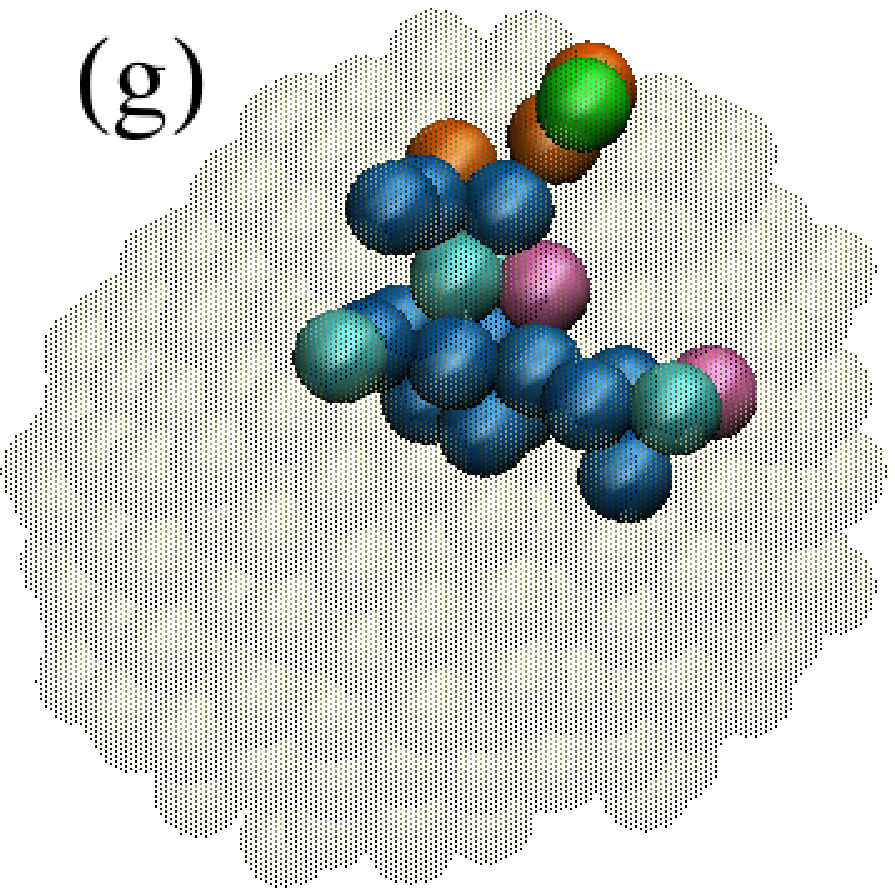}
\hspace{0.7cm}
\includegraphics[clip=true, trim=0 0 0 0, width=3.8cm]{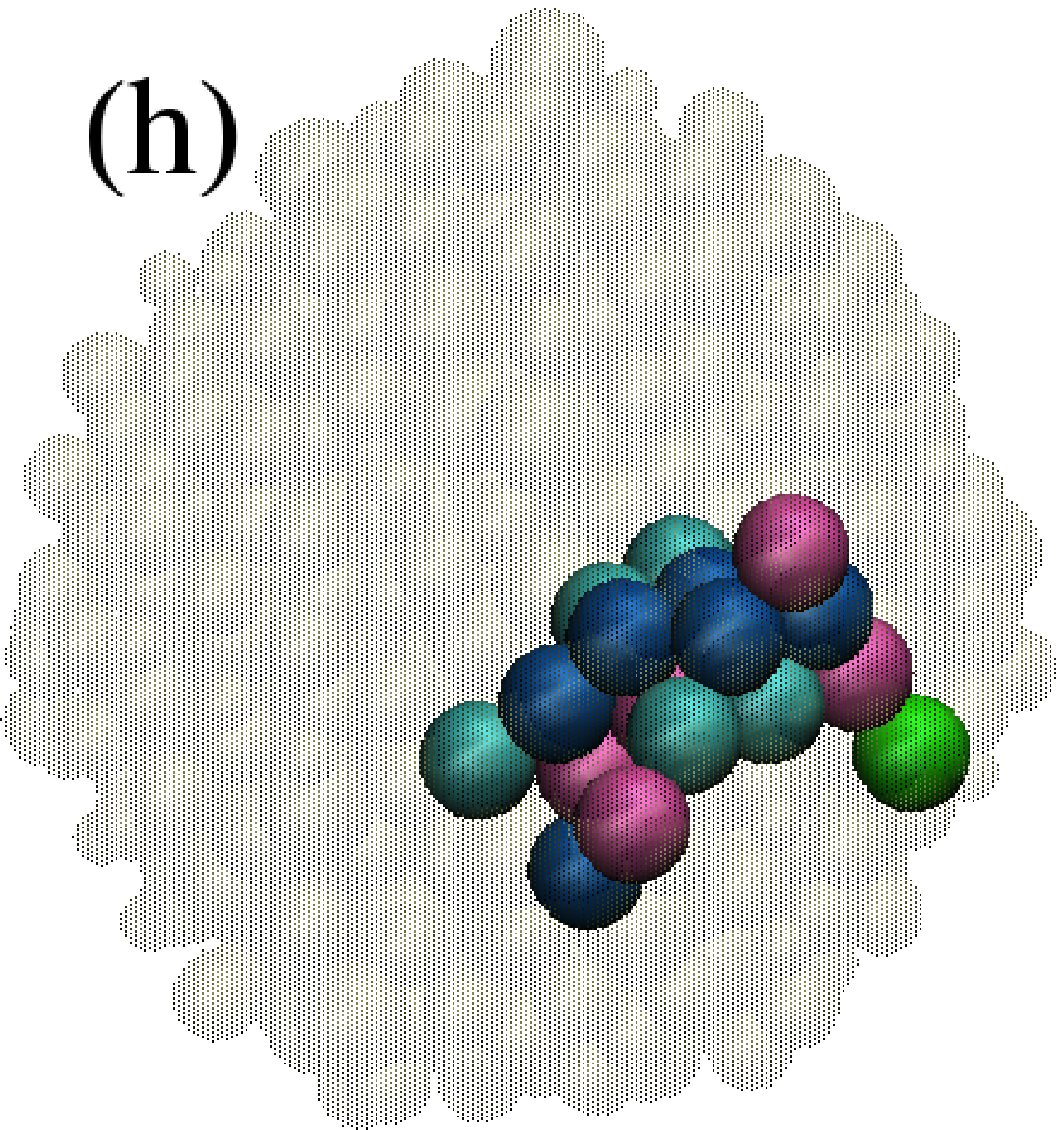}
\caption{Quenched droplet configurations containing embryos near critical size from 
(a) $T=0.485$, $Q_6=0.176$, $n_{\rm max}=98$, (b) $T=0.485$, $Q_6=0.086$, $n_{\rm max}=100$,
(c) $T=0.475$, $Q_6=0.129$, $n_{\rm max}=79$, (d) $T=0.475$, $Q_6=0.040$, $n_{\rm max}=76$,
(e) $T=0.465$, $Q_6=0.111$, $n_{\rm max}=63$, (f) $T=0.465$, $Q_6=0.038$, $n_{\rm max}=65$,
(g) $T=0.425$, $Q_6=0.078$, $n_{\rm max}=24$, (h) $T=0.200$, $Q_6=0.078$, $n_{\rm max}=18$,
The colouring scheme: blue, bulk fcc; mauve, bulk hcp;  yellow, bulk icosahedral; 
cyan, unidentified (amorphous); green, 111 surface; orange, 100 surface; particles not part of the critical embryo, transparent tan.
Note that the determination of the largest embryo is made prior to quenching and that the surface ordering visible for some 
of the droplets results from quenching.
}
\label{fig13}
\end{figure}

While we leave a more detailed study of these transformations near the critical region for the future, we show in
Fig.~\ref{fig13} a series of snapshots  of critical configurations from $T=0.485$ down to $T=0.200$. 
For $T=0.465$ and above, we select both high and low $Q_6$ specimens.  We assign particle types (fcc, hcp,
icosahedral) through common neighbour analysis~\cite{cna1,cna2}, which distinguishes between local structure
by considering the number of common neighbours two nearest neighbours share, as well as how those common neighbours are bonded.   Before carrying out the CNA analysis, we identify the particles in the largest embryo, and then carry out a conjugate
gradient quench of the system to remove vibrational displacements.  It is these quenched structures that are presented in
Fig.~\ref{fig13}, with particles originally in the largest embryo colour-coded, and the rest of the particles appearing
in a faint shade.

While we present here only a handful of structures, the picture that emerges seems rather robust.  The high $Q_6$
structures, Fig.~\ref{fig13}(a, c, e), appear to be stackings of fcc and hcp layers, while the low $Q_6$ structures appear 
to be multiply-twinned structures, rich in hcp, and possessing 5-fold symmetry.  For the lower $T$ shown, the embryos
are small and do not show secondary organization, but appear to be high in fcc.  Thus the embryos belonging to the incoming free energy trough in Fig.~\ref{fig12} appear to be randomly close-packed structures, while differentiation to structures suggestive of icosahedra or decahedra, occurs as or after these embryos approach critical size.

\begin{figure}[h]
\includegraphics[clip=true, trim=0 0 0 0, width=3.8cm]{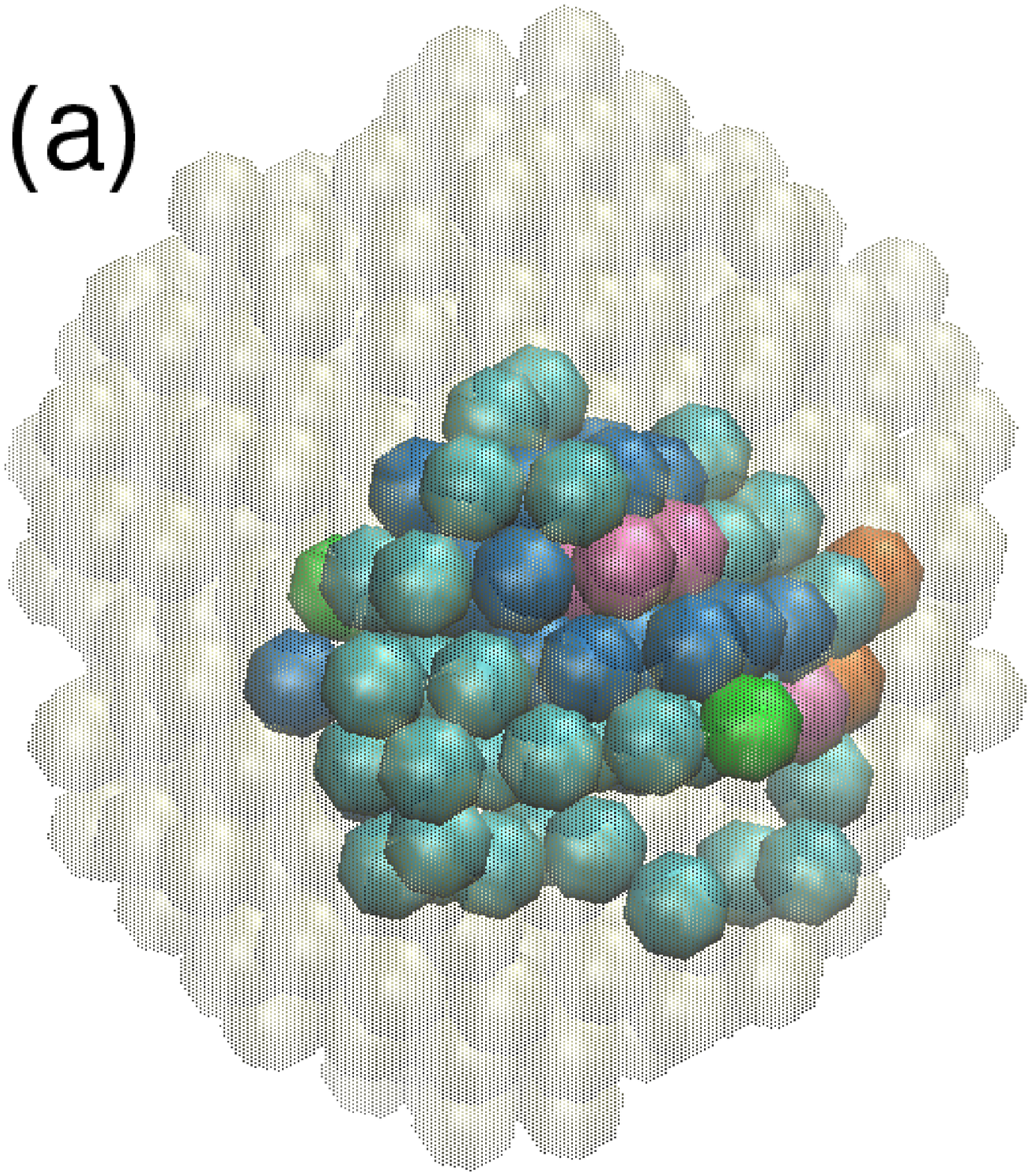}
\includegraphics[clip=true, trim=0 0 0 0, width=3.8cm]{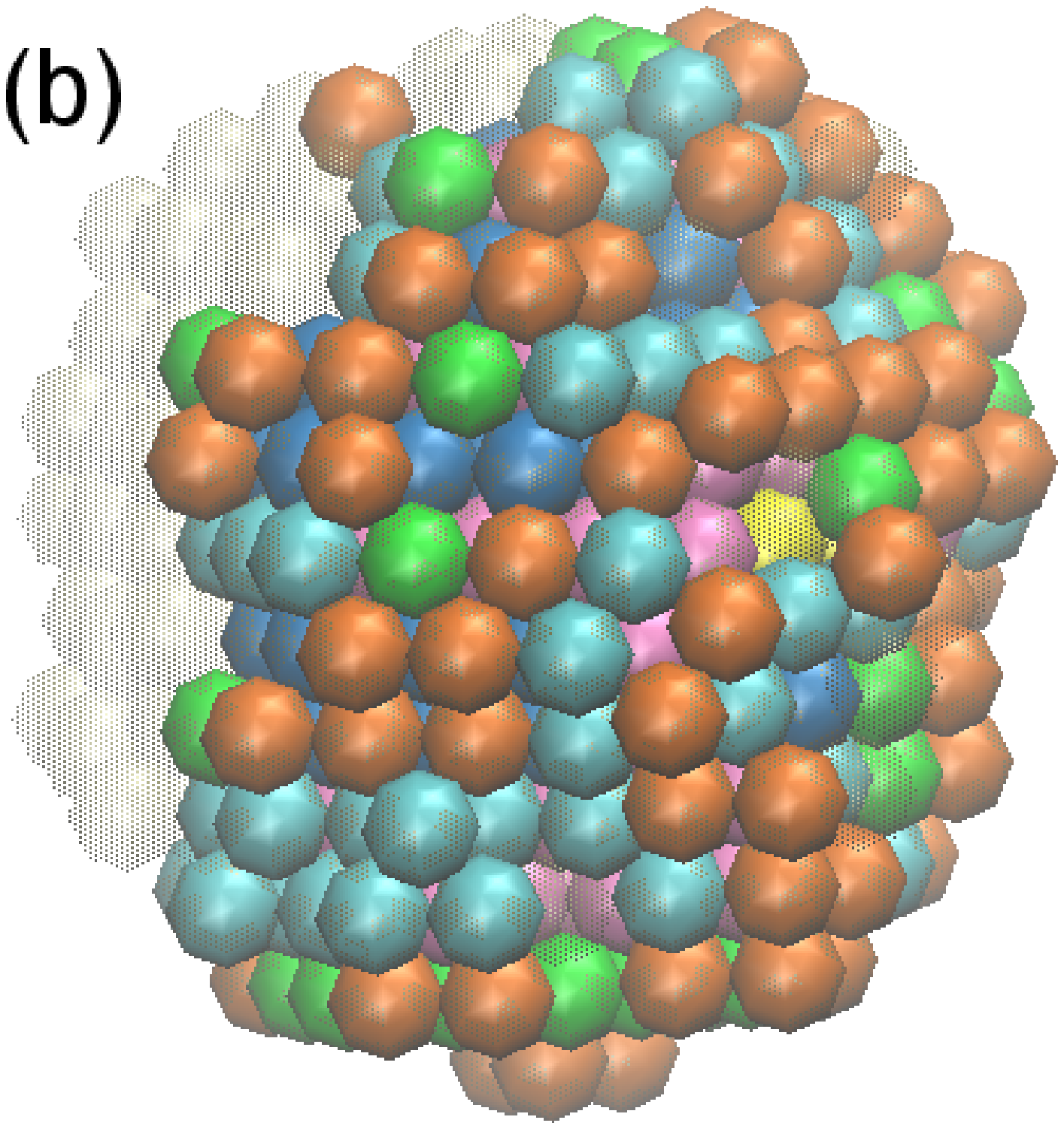}
\caption{
Effect of conjugate-gradient quench on droplet ordering.
Panel (a) shows the droplet from Fig.~\ref{fig13}(c) prior to quenching.
The surface is disordered and few particles are positively identified by CNA.
Panel (b) shows the same (quenched) droplet from Fig.~\ref{fig13}(c) but  with the determination of $n_{\rm max}$ done after quenching.  
Quenching induces significant ordering.
Color scheme is the same as in Fig.~\ref{fig13}.
}
\label{fig14}
\end{figure}

The reader may notice the significant ordering apparent on the surfaces of the clusters with larger embryos shown in 
Fig.~\ref{fig13}, e.g., the cluster in Fig.~\ref{fig13}(c). This ordering results from the conjugate-gradient quench performed to enhance the ability of CNA  to identify crystalline environments, and was previously noted in Ref.~\cite{TheRoleOfFcc}.  
Fig.~\ref{fig14}(a) shows the unquenched cluster at $T=0.475$ with CNA performed on the unquenched cluster as well.
In this case, there is significantly less surface ordering apparent and fewer particles within the embryo are assigned a particular classification.   
In Fig.~\ref{fig14}(b), we show that calculating $n_{\rm max}$ for the quenched cluster appearing in Fig.~13(c) results in many more solid-like particles, particularly on the surface. 
Thus, quenching presents a trade-off: it enhances identification of crystalline types, but induces significant ordering around  embryos. 
The significant ordering induced by quenching is certainly interesting, and is consistent with the ``prestructured surface cloud'' around crystallites
pointed out in Ref.~\cite{dellago}.

\section{Discussion}

Part of the motivation for this work comes from previous studies on the interpretation of 
$\beta \Delta F_{\rm min}^*$ approaching zero, its relation to nucleation rates and liquid metastability and the 
appropriateness of using the largest embryo in the system as an order parameter.
While previous work misidentified this barrier disappearance as a condition for a spinodal
(and the refutation of this pointed out its size dependence)~\cite{bagchi}, it clearly signalled some sort of limit to metastability.  
The recent scenario laid out in Ref.~\cite{RegueraCROSS}, 
namely that it signals unavoidable crystallization achieved through growth-limited nucleation, is supported
by our work.  We add that in our case, growth-limited nucleation proceeds in an out-of-equilibrium liquid.
By growth-limited nucleation we mean that with near certainty, somewhere in the system a critical nucleus
will form through  $\sim n^*$ consecutive particle additions, and so crystallization is controlled by the rate at which liquid-like
particles attach themselves to crystal-like ones.  This is what we see when we predict the rate through Eq.~\ref{eqCNTrate},
which matches MD rate determination for almost the entire range of $T$, as seen in Fig.~\ref{fig4rev}(b).
This growth-controlled nucleation mechanism, the onset of which is determined in part by the size
of the system, is likely important in crystallization occurring in other liquid droplet systems~\cite{FRANZESE,WATER}.


For our LJ clusters, where nucleation originates within the bulk, CNT as formulated for homogeneous nucleation for bulk liquids 
works quite well.  We see that a controlling factor, despite the presence of the surface, is the temperature $T_m$ at which $\Delta \mu=0$
in bulk systems, even though  in our finite-sized system the coexistence temperature 
$T_m^c$ is significantly lower.

We find that the simple modelling often used in CNT, such as constant $\gamma$, $\Delta H$ and $A$, and Arrhenius temperature
dependence of $f_{\rm crit}^+$ is supported by our results in independently determining these quantities,
at least for moderate supercooling.
Using these quantities as
calculated allows us to fit the rate with a single value of $\gamma=0.13$ convincingly well over a broad range of $T$.
There is  very good consistency between thermodynamics and rates, at least for $T\ge0.40$.
There is some ambiguity regarding the values of $\Delta H$, or rather $\Delta \mu$, and $\gamma$ when using the CNT model in 
Eq.~\ref{eqwork-min} to compare independently calculated $\beta \Delta G(n)$,
as the $\beta \Delta G(n)$ curves yield different values of $\Delta \mu$, and $\gamma$.  This points to the need for more
detailed modelling of $\beta \Delta G(n)$~\cite{SarahIvan,russo2014}.



The temperature $T_x=0.405$ at which system metastability is lost and growth-limited nucleation sets in is well
approximated by the condition $\beta\Delta G^*= -\ln{N_p}$~\cite{RegueraCROSS}. 
Near this same $T_x$, $f_{\rm crit}^+$ begins a rather strong departure from higher $T$ behavior by becoming  roughly constant.
This may indicate the liquid's inability to equilibrate because of sluggish dynamics, but may be at least partially 
driven by the weak $T$ dependence of $n^*$ that sets in below $T_x$.
That the system as a liquid does not reach metastable equilibrium is indicated by the potential energy time series at low $T$.
The ability for the liquid to undergo significant diffusive motion (enough to form critical embryos) while not equilibrating itself
may be due to a decoupling of diffusive and collective relaxation time scales characteristic of glassy dynamics~\cite{glassynuc}.
It is slightly curious that the onset of glassy dynamics should coincide with the system size dependent $T_x$.

As for the MC simulations, the constraint should allow for equilibration to occur since the size of the largest embryo is
constrained.  It is perhaps likely that relaxation of the metastable liquid requires significantly longer times  than our MC
of $500 000$ iterations ($5\times 10^6$ displacement attempts per particle).
Questions about the relaxation of the liquid
surrounding embryos are perhaps more easily addressed in bulk systems, where determining the dynamics of the system
is somewhat more straightforward in the absence of a surface.
While the increase in $n_{\rm inf}$ that we see at low $T$ may be viewed positively for the case of the spinodal scenario, the difficulties in discerning critical embryos precisely and questions regarding equilibrium must be carefully addressed.
Nonetheless, it is remarkable and slightly curious that the (more) equilibrated MC simulations at low $T$ 
should predict the rate so well through Eq.~\ref{eqCNTrate} when there is such a large difference in $n^*$ when comparing MD and MC.

Commenting on early work~\cite{TheRoleOfFcc}, where the free energy was calculated as a function of $Q_6$-based measures
of the bulk and surface crystallinity, at $T=0.475$ the barrier separating the liquid from a low $Q_6$ 5-fold structure was $0.5 k_{\rm B}T$ or less
(as calculated by subtracting from the free energy of the saddle point the minimum value in the liquid basin),
implying that the system as a liquid had (practically) lost stability at this $T$.  However, here we see that at $T=0.475$, $\beta \Delta F^*=10$,
which is considerably higher.  Thus, care must be taken when gauging phase stability from free energies based on $Q_6$, as there are crystal-like
states with values of $Q_6$ that overlap with those of the liquid.


In terms of structural differentiation that occurs in cluster crystallization, the picture that emerges in our work  
is that pre-critical nuclei are layered hcp-fcc planes, but that (at least) two types of structures, with different $Q_6$ values, leave the critical region. 
A small barrier in $Q_6$ appears to separate the two, thus preventing MD simulations from sampling the low $Q_6$ states  with twinned, five-fold structure until the embryo 
exceeds the critical size.  It seems that small icosahedral nuclei are unfavourable,
an observation that may find support in studies of small isolated LJ clusters~\cite{WALEDOYE}. 
The lack of sampling of low $Q_6$ critical states leads to disparity in determining $n^*$ in MD and MC, and will
make it more difficult to use the MFPT formalism to reconstruct the free energy landscape.
As nucleation studied here occurs within the bulk of the cluster,
perhaps a similar scenario occurs in bulk LJ.  We look forward to exploring these issues in more detail in the future.

\section{Conclusions}

We determine the rate of nucleation in a cluster of 600 LJ particles  through MD simulations 
by calculating mean first-passage times
of  embryo sizes.  For several orders of magnitude, the rate follows 
expectations from CNT under the simplest of
assumptions, namely a constant (ellipsoidal) shape of crystallites, a constant enthalpy difference, Arrhenius dependence
of the attachment rate, a melting temperature following from the bulk and a constant surface tension.  
Treating the surface tension as a fitting parameter to the rate while independently determining the other quantities results in excellent
agreement from $T=0.485$ down to $T\approx 0.40$ of the temperature dependence of the rate with CNT and of the work of forming critical nuclei with MC simulations.
However, the value of the effective surface tension $\gamma=0.13$ is smaller than expected.

Near  $T_x=0.4$, the rate starts approaching a maximum as the system loses its ability to maintain metastability.
This is evidenced by a monotonically
decreasing free energy that has as its argument the size of the largest embryo in the system.  At and below this temperature, crystallization proceeds through growth-limited
nucleation in an unequilibrated liquid.  The liquid phase is not inherently unstable itself, as there is a finite work required to form critical nuclei, but rather the barrier has become
sufficiently small, as determined approximately by $\beta \Delta G^* = \ln N_p$.  This picture follows what was observed for the vapour-liquid 
transition~\cite{RegueraCROSS}.

Surprisingly robust are the excellent predictions of the rate from MC-based calculations of $\beta \Delta G^*$, $Z$ and $f_{\rm crit}^+$.  The predictions
match the rate excellently above and below $T_x$.  Above $T_x$, the free energy $\beta \Delta F(n_{\rm max})$ gives the same barrier heights as $\beta \Delta G(n)$, given proper normalization.

For our system, MD and MC show discrepancies in $n_F^*$, even at slight to moderate supercooling, 
because of the appearance of embryos with twinned structures exhibiting 5-fold symmetry.  The differentiation between these and 
hcp-fcc stacked structures happens only in the critical region; pre-critical nuclei do not seem to possess the 5-fold symmetry of the icosahedral structures  to which LJ clusters often freeze.  In the critical region, there appears to be a small free energy barrier with $Q_6$ as an order parameter
between the hcp-fcc and 5-fold structures, inhibiting MD trajectories from sampling the same structures accessible to constrained MC simulations.

\section*{Acknowledgments}

We thank Richard K.~Bowles, Peter H.~Poole, Sergey V.~Buldyrev and especially David Reguera for warm and enlightening discussions.
We thank Natural Sciences and Engineering Research Council (Canada) for funding.
Computational facilities are provided by ACEnet, a member of Compute Canada and 
the regional high performance computing consortium for universities in Atlantic Canada. 
ACEnet is funded by the Canada Foundation for Innovation (CFI), 
the Atlantic Canada Opportunities Agency (ACOA), 
and the provinces of Newfoundland and Labrador, Nova Scotia, and New Brunswick.

\appendix
\section{Tabulated simulation results}

\begin{table}[H]
\begin{center}
\scalebox{0.82}{
    \begin{tabular}{| *{15}{c|}} 
    \hline
    \hline
    $\rm T$ & $\beta\Delta G^*$ & $\beta\Delta F^*$ & $\beta\Delta F^{*}_{{\rm min}}$ & $n^*$ & $n^*_{F}$ & $n^*_{\rm inf}$ & $Z$ & $f^+_{\rm crit}$ & $\rm S^*$ & $\rm J_{n^*} \times 10^{5}$\\
    \hline
    \hline
    0.485 & 18.80 & 18.80 & 11.08 & 100 & 100  & 94    & 0.0167  & 42.81    & 143.08   & 0.22\\
    0.480 & 17.69 & 17.69 & 9.93  & 89   & 89    & 87    & 0.0172  & 32.45    & 134.40   & 0.66\\
    0.475 & 16.69 & 16.69 & 8.89  & 79   & 77    & 96    & 0.0216  & 22.43    & 130.42   & 1.89\\
    0.470 & 15.76 & 15.76 & 7.92  & 71   & 73    & 89    & 0.0219  & 29.18    & 127.70   & 4.08\\
    0.465 & 14.88 & 14.88 & 6.99  & 62   & 62    & 88    &         &          &          & 7.75\\
    0.460 & 14.21 & 14.22 & 6.26  & 57   & 57    & 83    & 0.0249  & 20.67    & 113.82   & 15.87\\
    0.455 & 13.50 & 13.5   & 5.47  & 51   & 52    & 71    &         &          &          & 32.67\\
    0.450 & 12.97 & 12.89 & 4.79  & 46   & 48    & 66    & 0.0249  & 16.99    & 103.53   & 56.35\\
    0.445 & 12.29 & 12.29 & 4.10  & 41   & 42    & 56    &         &          &          & 98.86\\
    0.440 & 11.81 & 11.73 & 3.45  & 36   & 35     & 51    & 0.0256  & 11.96    & 88.61    & 149.73\\
    0.435 & 11.08 & 11.01 & 2.91  & 26   & 26     & 46    &         &          &          & 227.71\\
    0.430 & 10.75 & 10.60 & 2.31  & 24   & 23    & 41    & 0.0445  & 6.37     & 71.65    & 337.16\\
    0.425 & 10.45 & 10.13 & 1.65  & 20   & 18    & 34    &         &          &          & 486.30\\
    0.420 & 9.83   & 9.67   & 1.14  & 18   & 15    & 31    & 0.0553  & 3.59     & 57.91    & 647.61\\
    0.415 & 9.19   & 9.11   & 0.74  & 13   & 11    & 29    &         &          &          & 791.17\\ 
    0.410 & 8.76   & 8.71   & 0.46  & 13   & 9     & 27    & 0.0825  & 1.26     & 39.98    & 941.44\\
    0.405 & 8.26   & 8.20   & 0.14  & 10   & 7     & 26    &         &          &          & 1038.43\\
    0.400 & 7.96   &        &       & 9    &       & 25    & 0.1128  & 0.49     & 30.74    & 1275.26\\
    0.395 & 7.58   &        &       & 8    &       & 24    &         &          &          & 1366.98\\
    0.390 & 7.42   &        &       & 8    &       &       &         &          &          & \\
    0.385 & 7.22   &        &       & 7    &       & 23    &         &          &          & 1580.74\\
    0.380 & 7.07   &        &       & 7    &       &       & 0.1392  & 0.24     & 25.09    & \\
    0.375 & 7.09   &        &       & 7    &       &       &         &          &          & \\
    0.370 & 7.01   &        &       & 7    &       &       &         &          &          & \\
    0.350 & 6.86   &        &       & 6    &       & 18    & 0.1370  & 0.24     & 19.88    & 2705.77\\
    0.300 & 6.75   &        &       & 6    &       & 22    & 0.1350  & 0.16     & 20.81    & 2875.35\\
    0.250 & 5.90   &        &       & 4    &       & 32    &         &          & 12.17    & 1873.84\\
    0.200 & 6.03   &        &       & 4    &       & 59    &         &          & 14.02    & 734.11\\
    0.150 & 5.97   &        &       & 4    &       &       &         &          & 11.19    & \\
    \hline
    \end{tabular}
}
    \caption{
    Simulation results data.  For $\beta \Delta F^*$, we have added $\ln N_p = \ln 600 = 6.40$ in order to
    better compare with $\Delta G^*$.  For example, at $T=0.485$, the bare value of $\beta \Delta F^* = 12.40$ and at 
    $T=0.410$,  the bare value of $\beta \Delta F^* = 2.31$.}
\label{JFitPArameters}
\end{center}
\end{table}

We list detailed results in Table~\ref{JFitPArameters}: Barrier heights, $\beta\Delta G^*$ from MC, $\beta\Delta F^*$ from MC, $\beta\Delta F^{*}_{{\rm min}}$ from MC; critical sizes, $n^*$ from MC, $n^*_F$ from MC, $n^*_{\rm inf}$ from MD; Zeldovich factor $Z$ from MC; attachment rate $f^+_{\rm crit}$ based on embryos taken from MC; surface area $\rm S^*$ of critical embryos taken from MC; and nucleation rate $\rm J_{n^*}$ from MFPT data.

\end{document}